\documentclass[12pt]{article}

\usepackage{amsmath,amsfonts,enumerate,array}
\usepackage{graphicx}
\usepackage[usenames]{color}
\usepackage[english]{babel}
\usepackage{fancybox}
\usepackage{chngpage} 

\newcommand{\captionfonts}{\small}

\makeatletter  
\long\def\@makecaption#1#2{%
  \vskip\abovecaptionskip
  \sbox\@tempboxa{{\captionfonts #1: #2}}%
 \ifdim \wd\@tempboxa >\hsize
    {\captionfonts #1: #2\par}
  \else
    \hbox to\hsize{\hfil\box\@tempboxa\hfil}%
  \fi
  \vskip\belowcaptionskip}
\makeatother   

\usepackage{mciteplus}

\usepackage[colorlinks=true,      linkcolor=blue,      urlcolor=blue,      filecolor=blue,      citecolor=blue,
      pdfstartview=FitH,     pdfpagemode=UseNone,      bookmarksopen=true]{hyperref}  
      
\usepackage[all]{hypcap}     

\usepackage{cite}

\begin{document}

\numberwithin{equation}{section}


\mathchardef\mhyphen="2D


\newcommand{\be}{\begin{equation}} 
\newcommand{\ee}{\end{equation}} 
\newcommand{\bea}{\begin{eqnarray}\displaystyle}
\newcommand{\eea}{\end{eqnarray}}
\newcommand{\bt}{\begin{tabular}}
\newcommand{\et}{\end{tabular}}
\newcommand{\bs}{\begin{split}}
\newcommand{\es}{\end{split}}

\newcommand{\I}{\text{I}}
\newcommand{\II}{\text{II}}

\renewcommand{\a}{\alpha}	
\renewcommand{\b}{\beta}
\newcommand{\g}{\gamma}		
\newcommand{\G}{\Gamma}
\renewcommand{\d}{\delta}
\newcommand{\D}{\Delta}
\renewcommand{\c}{\chi}			
\newcommand{\C}{\Chi}
\newcommand{\p}{\psi}			
\renewcommand{\P}{\Psi}
\newcommand{\s}{\sigma}		
\renewcommand{\S}{\Sigma}
\renewcommand{\t}{\tau}		
\newcommand{\e}{\epsilon}
\newcommand{\n}{\nu}
\newcommand{\m}{\mu}
\renewcommand{\r}{\rho}
\renewcommand{\l}{\lambda}

\newcommand{\nn}{\nonumber\\} 		
\newcommand{\newotimes}{}  				
\newcommand{\diff}{\,\text{d}}		
\newcommand{\h}{{1\over2}}				
\newcommand{\Gf}[1]{\G \Big{(} #1 \Big{)}}	
\newcommand{\floor}[1]{\left\lfloor #1 \big\rfloor}
\newcommand{\ceil}[1]{\left\lceil #1 \right\rceil}

\def\c{\color{blue}}

\def\b{\bigskip}

\begin{flushright}
\end{flushright}
\vspace{20mm}
\begin{center}
{\LARGE Bootstrapping multi-wound twist effects\\\vspace{2mm} in symmetric orbifold CFTs}
\\
\vspace{18mm}
\textbf{Bin} \textbf{Guo}{\footnote{bin.guo@ipht.fr}} ~ \textbf{and} ~ \textbf{Shaun}~ \textbf{D.}  \textbf{Hampton}{\footnote{shaun.hampton@ipht.fr}}
\\
\vspace{10mm}

${}$Institut de Physique Th\'eorique,\\
	Universit\'e Paris-Saclay,
	CNRS, CEA, \\ 	Orme des Merisiers,\\ Gif-sur-Yvette, 91191 CEDEX, France  \\

\vspace{8mm}
\end{center}

\vspace{4mm}

\thispagestyle{empty}

\begin{abstract}
\noindent
We investigate the effects of the twist-2 operator in 2D symmetric orbifold CFTs.
The twist operator can join together a twist-$M$ state and a twist-$N$ state, creating a twist-$(M+N)$ state. This process involves three effects: pair creation, propagation, and contraction. We study these effects by using a Bogoliubov ansatz and conformal symmetry. In this multi-wound scenario, pair creation no longer decouples from propagation, in contrast to the previous study where $M=N=1$. 
We derive equations for these effects, which organize themselves into recursion relations and constraints.
Using the recursion relations, we can determine the infinite number of coefficients in the effects through a finite number of inputs. Moreover, the number of required inputs can be further reduced by applying constraints.

\vspace{3mm}

\end{abstract}
\newpage

\setcounter{page}{1}

\numberwithin{equation}{section} 

\tableofcontents

\newpage

\section{Introduction}

Symmetric orbifold CFTs have proven to be a useful class of theories in studying ${\rm AdS}_3/{\rm CFT}_2$ \cite{Maldacena:1997re,Strominger:1996sh,Maldacena:1999bp,Seiberg:1999xz,Dijkgraaf:1998gf,Larsen:1999uk,Jevicki:1998bm,deBoer:1998kjm,Gomis:2002qi,Gava:2002xb}. Among these, the symmetric orbifold CFT with $\mathcal N = (4,4)$ supersymmetry has played a prominent role in understanding the D1D5 system, both in the context of the free theory and beyond. Recent studies found that the free D1D5 CFT is dual to tensionless string theory on ${\rm AdS}_3$ \cite{Eberhardt:2018ouy,Eberhardt:2019ywk,Eberhardt:2020akk, Dei:2020zui,Knighton:2020kuh,Gaberdiel:2021kkp,Eberhardt:2021vsx}. The free theory also provides a microscopic count of black hole microstates from a field theory perspective \cite{Strominger:1996sh,Maldacena:1999bp}. In certain cases, these microstates are explicitly known \cite{Bena:2022ldq,Bena:2022rna,Shigemori:2020yuo}, corresponding to specific states in the CFT. Perturbing away from the free point has also been studied to address questions at the gravity point such as the lifting of energies \cite{Gaberdiel:2015uca,Hampton:2018ygz,Guo:2019pzk,Guo:2019ady,Guo:2020gxm,Lima:2020boh,Lima:2020nnx,Lima:2020urq,Lima:2020kek,Lima:2021wrz,AlvesLima:2022elo,Benjamin:2021zkn,Guo:2022ifr,Fiset:2022erp}, tidal scrambling \cite{Guo:2021ybz,Guo:2021gqd}, thermalization \cite{Hampton:2019hya,Hampton:2019csz}, and dynamical evolution of black hole microstates \cite{Guo:2022and}, etc. There has also been work exploring symmetric orbifolds of minimal models \cite{Belin:2020nmp,Apolo:2022fya,Benjamin:2022jin}.

Symmetric orbifold CFTs have the target space ${\mathcal M^{N}/S_N}$, where $\mathcal M$ represents the target space of the `seed' CFT, $N$ denotes the number of seed CFT copies, and $S_N$ is the permutation group. Due to the orbifold, there exist twist-$n$ operators $\sigma_n$, around which $n$ copies of the seed CFT permute into each other.
The traditional
way to compute correlation functions involving twist operators is to map the 2-d base space to
the corresponding covering space, as developed in \cite{Lunin:2000yv, Lunin:2001pw,Pakman:2009zz,Pakman:2009ab}. However, as the number of twist operators or copies involved increases, finding the covering map becomes increasingly difficult as it requires solving higher-order polynomial equations.

Recently a new method was developed without using the covering space \cite{Guo:2022sos,Guo:2022zpn}. It uses conformal symmetry and the nature of the twist operator as a Bogoliubov transformation. 
Specifically, the effect of the twist operator $\sigma_2$ joining two singly wound (untwisted) copies was computed.
Some early attempts in this direction can be found in \cite{Burrington:2014yia,Carson:2014xwa,Carson:2017byr}. For other related works that compute correlation functions of twist operators using conformal symmetry, see e.g. \cite{Dixon:1986qv,Arutyunov:1997gt,Arutyunov:1997gi,Jevicki:1998bm,Dei:2019iym}. 
The potential generalization of this method to correlation functions where a covering map is not applicable makes it interesting. Furthermore, the perturbation of the CFT away from the free point involves the twist operator $\sigma_2$. Computing the effect of this operator is also crucial for understanding the physics at the gravity point. 
In this work, we generalize this method to compute the effect of the twist operator $\sigma_2$ on 
a twist-$M$ state and a twist-$N$ state, creating a twist-$(M+N)$ state.
This process involves $M+N$ copies of the seed CFT, which provides a valuable test of the method when more copies are involved.

There are three effects of the twist operator $\sigma_2$: pair creation, propagation, and contraction. 
The main challenge in this work lies in the fact that pair creation is no longer independently determined from propagation; instead, they are coupled together algebraically. This complication arises from the fact that when $L_{-1}$ acts on a twisted vacuum, it does not yield zero as it does when acting on an untwisted vacua. To solve this coupled system, we organize the equations into recursion relations and constraints. Using the recursion relations, we can determine the infinite number of coefficients in the effects through a finite number of inputs. The number of required inputs can be further reduced by applying constraints. 

The plan of this paper is as follows. In section \ref{section 2}, we review the symmetric orbifold CFT of a single free boson. In section \ref{section 3}, we review the effects of the twist operator. In section \ref{section 4}, we discuss the Bogoliubov transformation. In section \ref{section 5}, we derive relations for the effects of the twist operator by using conformal generators and the Bogoliubov ansatz. In section \ref{section 6}, we summarize our results, organizing the relevant equations into recursion relations and constraints. In section \ref{section 7}, we briefly recall the results of the twist effects derived using covering maps for the multi-wound scenario considered in this paper. In section \ref{section 8}, we provide specific examples for some values of $M$ and $N$. In section \ref{section 10}, we discuss our results and future directions.

\section{Symmetric orbifold CFT for one free boson}\label{section 2}

Symmetric orbifold CFTs are obtained by orbifolding $N$ copies of a seed CFT with target space $\mathcal M$ by the permutation group $S_N$, which yields the target space
\be
\mathcal M^N / S_N
\ee
In this paper, we study a seed CFT of one free boson, where $\mathcal M = \mathbb{R}$ and $c=1$. Extension to theories containing multiple free bosons and fermions can be obtained similarly \cite{Guo:2022zpn}.
This theory contains untwisted sectors and twisted sectors. Consider a single bosonic field $X(w)$ living on a cylinder defined by the coordinate
\be
w =\tau + i\sigma,\quad  -\infty < \tau < \infty,\quad 0\leq \sigma < 2\pi
\ee
Consider $N$ such copies of this boson
\bea 
X^{(i)}(w),\quad 1\leq i\leq N 
\eea
Pick $k$ of the $N$ copies and define the $k$ twisted sector by the action of a twist operator of order $k$, $\s_k$. What this action does is change the boundary conditions of the bosonic fields $X^{(i)}$ as one takes $\s\to \s +2\pi$ as follows
\bea
X^{(i)}(\s + 2\pi)&=&X^{(i+1)}(\s ),\quad 1\leq  i \leq k-1
\nn 
X^{(k)}(\s+2\pi)&=&X^{(1)}(\s)
\eea
This defines the $k$ twisted sector in the orbifold theory. The remaining copies are in the untwisted sector and are defined by
\be
X^{(j)}(\s+2\pi)=X^{(j)}(\s),\quad k+1\leq j\leq N 
\ee
Oscillator modes can be defined in the singly wound (untwisted) sector for copy $i$, at time $\tau$, as 
\be
\a^{(i)}_{m}={1\over2\pi}\int_{\sigma=0}^{2\pi} dw\, e^{mw}\partial X^{(i)}(w),\quad m\in \mathbb{Z} 
\ee
The commutation relations are 
\bea
[ \a^{(i)}_{m},\a^{(j)}_{n} ]=m\d^{ij}\d_{m+n,0}
\eea
The bosonic modes obey the reality condition
\bea
\a^{(i)\dagger}_{m} = \a^{(i)}_{-m}
\eea
and the bosonic vacuum is defined by the condition
\bea 
\a^{(i)}_{m}|0\rangle^{(i)} = 0, \quad m\geq 0
\eea
The theory also contains Virasoro generators $L_{n}$ with $n\in \mathbb{Z}$ which can be written as 
\bea \label{L single}
L_{n}={1\over2}\sum_i\sum_m\a^{(i)}_{m}\a^{(i)}_{n-m},\quad m,n\in \mathbb{Z}
\eea
Their commutation relations with bosonic modes of copy $i$ are given by
\bea\label{L1 singly wound} 
[L_n,\a^{(i)}_{m}] = -m\a^{(i)}_{m+n}
\eea
Oscillator modes can also be defined for a $k$-wound (twist-$k$) bosonic field for a constant time $\t$ as follows
\bea
\a_{m}={1\over2\pi}\int_{\sigma=0}^{2\pi k} dw \, e^{mw}\partial X(w),\quad m = {p\over k},\quad p\in \mathbb{Z}
\eea 
In the $k$ twisted sector the vacuum is defined by
\bea 
\a_{m}|0_k\rangle = 0, \quad m\geq 0
\eea
where the $k$ twisted vacuum $|0_k\rangle$ can be created by the twist-$k$ operator $\sigma_k$ in the following manner 
\bea\label{twisted vacuum}
|0_k\rangle \equiv \s_k(w\to-\infty)\prod_{i=1}^{k}|0\rangle^{(i)}
\eea
We will also use the term `$k$ wound' to denote `twist-$k$'.
The commutation relation for the modes are given by
\bea
[ \a_{m},\a_{n} ]=km\d_{m+n,0}
\eea
The Virasoro generators in the $k$ twisted sector take the form.
\bea\label{L multi} 
L_n = {1\over2k}\sum_m\a_m\a_{n-m},
 \quad m, n = {p\over k},\quad  p \in \mathbb{Z}
\eea
The commutation relation between the Virasoro generators and a single bosonic mode is still given by
\bea\label{L alpha multiwound}
[L_n,\a_{m}]= - m\a_{m+n}
\eea
We note that this relation has the same form as (\ref{L1 singly wound}). The dimension of the twist-$k$ operator is given by
\bea\label{sig k}   
 h(\s_k) = {1\over24}\bigg( k-{1\over k} \bigg)
\eea

\section{Effects of a twist operator}\label{section 3}

In this section we discuss different effects of the twist operator on the bosonic modes. In this paper we only consider the twist-2 operator $\sigma_2$ and we thus define
\bea
\s\equiv\s_2
\eea
which has dimension
\bea
h\equiv h_2 = {1\over16}
\eea
In addition, we consider processes where an $M$ wound copy and an $N$ wound copy in the initial state are twisted into an $M+N$ wound copy. As such let's consider a state of the following form
\bea
\a^{(i_1)}_{-{m_1\over k_1}}\a^{(i_2)}_{-{m_2\over k_2}}\ldots\a^{(i_n)}_{-{m_n\over k_n}}|0_M\rangle^{(1)}|0_N\rangle^{(2)} 
\eea
where $m_j$ are positive integers and $k_j = M$ for $i_j = 1$, and $k_j = N$ for  $i_j = 2$.
Now, consider the action of a single twist on this state
\bea
\s(w)\,\a^{(i_1)}_{-{m_1\over k_1}}\a^{(i_2)}_{-{m_2\over k_2}}\ldots\a^{(i_n)}_{-{m_n\over k_n}}|0_M\rangle^{(1)}|0_N\rangle^{(2)} 
\eea
This will produce a final state in the $M+N$ wound sector. We would like to study the effects which characterize this state. These effects come in three types

\b

(i) Contraction:
The first effect is called contraction where any two modes, $\a^{(i)}_{-{r\over k_i}}\a^{(j)}_{-{s\over k_j}}$ in the initial state (before the twist) contract together. We denote this by
\bea
C[\a^{(i)}_{-{r\over k_i}},\a^{(j)}_{-{s\over k_j}}] \equiv C^{ij}_{{r\over k_i},{s\over k_j}}
\eea
We consider all possible pairs of contractions. If the modes don't contract then we must consider another process where each mode moves across the twist to form a linear combination of modes in the final state. We consider this in step (ii).

\b

(ii) Propagation: When a mode passes through the twist it produces a linear combination of modes in the final state. We call this propagation, and the process is characterized by the functions 
\bea
f^{(i)}_{{r\over k_i},{p\over M+N}},\quad i=1,2, \quad k_1=M,~~ k_2=N
\eea
where the first index labels the initial energy and the second index labels the final energy. 
We note that if the final mode numbers correspond to integer energies and they don't equal the initial energy then the propagation can be argued to vanish \cite{Guo:2022sos,Carson:2014xwa}  
\bea\label{f n0}
f^{(i)}_{{r\over k_i},{p\over M+N}} = 0,\quad \quad {r\over k_i}\neq {p\over M+N} \in \mathbb{Z}
\eea

\b

(iii) Pair creation: The third effect comes when the vacuum itself is twisted together. We write this effect as the following
\bea\label{chi}
|\chi\rangle &\equiv& \s(w)|0_M\rangle^{(1)}|0_N\rangle^{(2)} \cr 
&=&A\,\text{exp}\Big(\sum_{p,q>0}\gamma_{{p\over M+N},{q\over M+N}}\a_{-{p\over M+N}}\a_{-{q\over M+N}}\Big)|0_{M+N}\rangle
\eea
where 
\bea\label{g n0}
\gamma_{{p\over M+N},{q\over M+N}}\neq 0,\quad {p\over M+N},{q\over M+N}\notin \mathbb{Z}
\eea
$A$ is a function that captures the contributions of the vacuum correlations of just the twist operators themselves. We will not need to determine this function explicitly as it will be divided out in our computations. 

To be more precise and illustrative about these effects, let's consider a scenario involving only a single mode in the initial state. As there are no other modes present, this mode cannot contract with them and thus, must propagate through the twist operator. Additionally, there is also the pair creation effect. The final result is
\bea\label{f M}
\s(w)\a^{(i)}_{-{r\over k_i}}|0_M\rangle^{(1)}|0_N\rangle^{(2)} = \sum_{p>0}f^{(i)}_{{r\over k_i},{p\over M+N}}\a_{-{p\over M+N}}|\chi\rangle
\eea
Let's consider two modes in the initial state. The state produced by the twist operator is given by
\bea\label{copy 1 copy 1}
&&\s(w)\a^{(i)}_{-{r\over k_i}}\a^{(j)}_{-{s\over k_j}}|0_M\rangle^{(1)}|0_N\rangle^{(2)}\nn 
&=&\Big(C^{ij}_{{r\over k_i},{s\over k_j}}+ \sum_{p>0}f^{(i)}_{{r\over k_i},{p\over M+N}}\a_{-{p\over M+N}}\sum_{q>0}f^{(j)}_{{s\over k_j},{q\over M+N}}\a_{-{q\over M+N}}\Big)|\chi\rangle
\eea
The first term arises from the contraction of the two initial modes. The second term comes from the propagation of the two modes through the twist operator. The state $\chi$ captures the pair creation effect. 
In the next section, we explain the motivation behind the ansatz for these effects.

\section{Bogoliubov Transformation}\label{section 4}

Here we explain the Bogoliubov transformation and how it motivates the ansatz for the effects of the twist. We use a simple example. Consider a set of creation and annihilation operators $\hat a,\hat a^{\dagger}$ of some quantum theory. The vacuum of this theory is defined by 
\bea
\hat a|0\rangle_a=0
\eea
Now consider a change of basis in which the original operators are written in terms of a new set $\hat b,\hat b^\dagger$ with a new vacuum state defined by
\bea
\hat b|0\rangle_b=0
\eea
We can write the operators $\hat a,\hat a^{\dagger}$ as a linear combination of the operators $\hat b,\hat b^{\dagger}$ as follows
\bea
\hat a &=& \alpha\, \hat b + \beta\, \hat b^{\dagger}\cr 
\hat a^{\dagger} &=& \alpha^*\, \hat b^{\dagger} + \beta^*\, \hat b
\eea
where $|\a|^2-|\beta|^2=1$ to make the commutation relations canonical
\bea 
[\hat a,\hat a^{\dagger}]=1,\qquad [\hat b,\hat b^{\dagger}]=1 
\eea
We can now see that $\hat b$ does not annihilate the $a$ vacuum, i.e.
\bea 
\hat a|0\rangle_a=0\implies (\alpha\, \hat b + \beta\,\hat b^{\dagger})|0\rangle_a=0
\eea
This equation is solved if
\be\label{a vac}
|0\rangle_a = e^{{1\over2}\g\, \hat{b}^{\dagger}\hat{b}^{\dagger}}|0\rangle_b
\ee
where 
\be\label{g}
\g = -  {\beta\over \a}
\ee
This shows that the vacuum in terms of one set of modes can be an excited state with respect to another set of modes. 
Let's consider an example to further demonstrate our motivation. Consider two excitations on the `$a$' vacuum
\bea 
\hat a^{\dagger}\hat a^{\dagger}|0\rangle_a
\eea
Rewriting this in terms $\hat b,\hat b^{\dagger}$ we have
\bea
\hat a^{\dagger}\hat a^{\dagger}|0\rangle_a= (\alpha^{*} \hat b^{\dagger} + \beta^{*} \hat b)(\alpha^{*} \hat b^{\dagger} + \beta^{*} \hat b)|0\rangle_a
\eea
Using commutation relations and (\ref{a vac}), this can be rearranged as
\bea 
\hat a^{\dagger}\hat a^{\dagger}|0\rangle_a = \big((\a^* + \beta^*\g)\hat b^{\dagger}(\a^* + \beta^*\g)\hat b^{\dagger} + \beta^*\a^*\big) e^{{1\over2}\g\, \hat b^{\dagger}\hat b^{\dagger}}|0\rangle_b
\eea
We can see the similarity between this and our ansatz for the effects of the twist operator where here the following effects can be identified:

\b 

(i) Contraction is identified with the term
\bea 
C[\hat a^{\dagger}\hat a^{\dagger}]=\beta^*\alpha^*
\eea

\b

(ii) Propagation is given by the term
\bea 
\hat a^{\dagger}\to (\a^* + \beta^*\g)\hat b^{\dagger}
\eea

\b 

(iii) Pair creation
\bea
|0\rangle_a = e^{{1\over2}\g \hat b^{\dagger}\hat b^{\dagger}}|0\rangle_b
\eea
These coefficients are not independent but are related through (\ref{g}). We will call these three rules, along with the constraint (\ref{g}), the `normal' Bogoliubov ansatz. 

In \cite{Carson:2014xwa}, the effect of the twist operator was studied using a Bogoliubov transformation along with a constraint similar to the one in (\ref{g}) where the coefficients $\a,\beta,\g$ were promoted to infinite dimensional matrices $\a_{ij},\beta_{ij},\g_{ij}$. In this method the relation (\ref{g}) becomes a matrix multiplication which requires the inversion of infinite dimensional matrices. In this paper, following \cite{Guo:2022sos,Guo:2022zpn}, we use a `weak' Bogoliubov ansatz in which we use the ansatz outlined in section \ref{section 3}, without the matrix version of the constraint (\ref{g}). In addition, we instead use conformal symmetry in an attempt to compute the relative quantities. 

In the next section, using conformal generators, we find a set of expressions for pair creation and propagation.

\section{Bootstrapping effects of the twist operator in the multiwound sector}\label{section 5}

In this section, we use conformal symmetry and the Bogoliubov ansatz to derive equations for computing the effects of the twist operator in the multiwound scenario. The results are summarized in section {\ref{section 6}}.

\subsection{Relations using $L_0$}

Let us begin with relations that yield the $w$ dependence of the pair creation $\gamma$ and propagation $f$. 
To determine the $w$ dependence for $\gamma$, we start with the following equation
\bea
0 = \s(w)(L_0-h_M-h_N)|0_M\rangle^{(1)}|0_N\rangle^{(2)}
\eea
where we have used the fact that
\bea
L_0|0_P\rangle = h_P|0_P\rangle
\eea
where $h_P$ is defined in (\ref{sig k}), and the twisted vacuum is defined in (\ref{twisted vacuum}). By commuting $L_0$ through $\s(w)$, we obtain the relation
\bea\label{L0 comm 0}
0 = \big((L_0-h_M-h_N)\s(w) - [L_0,\s(w)]\big)|0_M\rangle^{(1)}|0_N\rangle^{(2)}
\eea
Let us compute the commutator in the second term. 
Since similar commutators appear frequently later on, we will derive the commutator for a general $L_n$
\bea \label{L comm}
[L_{n},\s(w)] &=& {1\over2\pi i}\oint dw' e^{nw'}T(w')\s(w)\cr  
&=&e^{nw}{1\over2\pi i}\oint dw'e^{n(w'-w)} T(w')\s(w)\cr 
&=&e^{nw}{1\over2\pi i}\oint dw'(1+n(w'-w)+\ldots) T(w')\s(w)\cr   
&=& e^{nw}(L^{(w)}_{-1}+nL^{(w)}_{0})\s(w)\cr   
&=& e^{nw}(\partial+n h)\s(w)
\eea 
Here, the modes with the superscript $(w)$, i.e., $L^{(w)}_{n}$, refer to those centered around $w$.
We have used the fact that $\sigma(w)$ is a primary of dimension $h=\frac{1}{16}$, satisfying the following conditions
\be\label{conditions}
L^{(w)}_{-1} \sigma(w) = \partial \sigma(w), \quad L^{(w)}_{0} \sigma(w) = h \sigma(w), 
\quad L^{(w)}_{n>0} \sigma(w) = 0
\ee
Inserting this into (\ref{L0 comm 0}) and using the ansatz (\ref{chi}) gives
\bea
0 &=& (L_0 - \partial -h_M-h_N)\s(w)|0_M\rangle^{(1)}|0_N\rangle^{(2)}\nn
 &=& (L_0 - \partial -h_M-h_N )A\exp\bigg(\sum_{p,q>0}\gamma_{\frac{p}{M+N},\frac{q}{M+N}}\a_{-\frac{p}{M+N}}\a_{-\frac{q}{M+N}}\bigg)|0_{M+N}\rangle
\eea
Expanding the exponential and looking at terms without any modes, we find the relation
\bea\label{A}
\partial A=(h_{M+N} - h_M-h_N)A
\eea
Considering terms with two modes $\a_{-\frac{p}{M+N}}\a_{-\frac{q}{M+N}}$, we obtain
\bea
0 =\left (\frac{p+q}{M+N} + h_{M+N} - \partial- h_M-h_N\right)A\,\gamma_{\frac{p}{M+N},\frac{q}{M+N}} 
\eea
By inserting the relation (\ref{A}), we find the following relation
\bea\label{gen one twist exp}
\left(\frac{p+q}{M+N} -\partial \right)\gamma_{\frac{p}{M+N},\frac{q}{M+N}} = 0
\eea 
This indicates that the $w$ dependence of $\gamma$ is given by
\bea\label{gam w dep M N}
\gamma_{\frac{p}{M+N},\frac{q}{M+N}}\propto e^{\frac{p+q}{M+N}w}
\eea

Next, Let us derive the $w$ dependence of the propagation $f$ using $L_0$.
We first look at copy $1$ and the results are easily extendable to copy $2$. We start with the relation
\bea
0 =\s(w)\left(L_0 -\big(h_M + h_N + \frac{r}{M} \big)\right)\a^{(1)}_{-\frac{r}{M}}|0_M\rangle^{(1)}|0_N\rangle^{(2)}
\eea
where we have used the commutation relation (\ref{L alpha multiwound}).
We can commute $L_0$ through the twist and then use (\ref{L comm}) to obtain the relation
\bea 
0= \left(L_0 - \partial - \big(h_M + h_N + \frac{r}{M} \big) \right)\s(w)\a^{(1)}_{-\frac{r}{M}}|0_M\rangle^{(1)}|0_N\rangle^{(2)}
\eea
Using the ansatz (\ref{f M}), we find
\bea
0&=& \left(L_0 - \partial - \big(h_M + h_N + \frac{r}{M} \big) \right)\nn
&&~~\sum_{p'> 0}f^{(1)}_{\frac{r}{M},\frac{p'}{M+N}}\a_{-\frac{p'}{M+N}}
A\exp\bigg(\sum_{p,q>0}\gamma_{\frac{p}{M+N},\frac{q}{M+N}}\a_{-\frac{p}{M+N}}\a_{-\frac{q}{M+N}}\bigg)|0_{M+N}\rangle
\eea
Looking at terms with one mode $\a_{-\frac{p}{M+N}}$, we obtain
\bea
0= \left(\big(h_{M+N} + \frac{p}{M+N} \big) - \partial - \big(h_M + h_N + \frac{r}{M} \big) \right) A \, f^{(1)}_{\frac{r}{M},\frac{p}{M+N}}
\eea
Inserting the expression (\ref{A}) to eliminate $A$ yields
\bea
 \left( \frac{p}{M+N} - \partial -  \frac{r}{M}  \right) f^{(1)}_{\frac{r}{M},\frac{p}{M+N}} = 0
\eea
Hence, we find that the $w$ dependence of the propagation $f^{(1)}$ is given by
\bea \label{f dep w}
f^{(1)}_{\frac{r}{M},\frac{p}{M+N}}\propto e^{\left(\frac{p}{M+N} - \frac{r}{M}\right)w}
\eea
We see that the exponential factor is the difference between the initial and final energies. 
Similarly, we can derive $f^{(2)}$ by interchanging $M$ and $N$ in the expression for $f^{(1)}$. Thus, we have
\bea \label{f dep w 2}
f^{(2)}_{\frac{r}{N},\frac{p}{M+N}}\propto e^{\left(\frac{p}{M+N} - \frac{r}{N}\right)w}
\eea

\subsection{Relations using $L_{-1}$}

Now let us derive relations using the conformal generator $L_{-1}$. We can write it as a sum of modes acting on each copy
\bea\label{L-1 copies}
L_{-1} = {1\over2M}\sum_{r}\a^{(1)}_{-\frac{r}{M}}\a^{(1)}_{-1+\frac{r}{M}} + {1\over2N}\sum_{r}\a^{(2)}_{-\frac{r}{N}}\a^{(2)}_{-1+\frac{r}{N}}
\eea
It is worth noting that $L_{-1}$ acting on the multiwound vacuum $|0_M\rangle^{(1)} |0_N\rangle^{(2)}$ is not zero, except in the case where $M=N=1$. This leads to nonlinear recursion relations that involve both pair creation $\gamma$ and propagation $f$. In contrast, when $M=N=1$ (as studied in \cite{Guo:2022sos,Guo:2022zpn}), the linear recursion relations only involve pair creation $\gamma$ and can be solved more easily. 
Using (\ref{L-1 copies}), we can write the following equation
\bea
0= \s(w)\bigg(L_{-1} - {1\over2M}\sum_{r=1}^{M-1}\a^{(1)}_{-\frac{r}{M}}\a^{(1)}_{-1+\frac{r}{M}} - {1\over2N}\sum_{r=1}^{N-1}\a^{(2)}_{-\frac{r}{N}}\a^{(2)}_{-1+\frac{r}{N}} \bigg)|0_M\rangle^{(1)} |0_N\rangle^{(2)}
\eea 
By commuting $L_{-1}$ through the twist and using (\ref{L comm}), we obtain
\bea\label{Lm1} 
0&=& \bigg(L_{-1}\s(w) - e^{-w}(\partial-h)\s(w)\nn 
&&- \s(w){1\over2M}\sum_{r=1}^{M-1}\a^{(1)}_{-\frac{r}{M}}\a^{(1)}_{-1+\frac{r}{M}} - \s(w){1\over2N}\sum_{r=1}^{N-1}\a^{(2)}_{-\frac{r}{N}}\a^{(2)}_{-1+\frac{r}{N}} \bigg)|0_M\rangle^{(1)} |0_N\rangle^{(2)}
\eea
Using (\ref{copy 1 copy 1}), we obtain
\bea \label{Lm1 2}
0&=&\bigg( L_{-1} - e^{-w}(\partial-h)-{1\over2M}\sum_{r=1}^{M-1} C^{11}_{\frac{r}{M},1-\frac{r}{M}}-{1\over2N}\sum_{r=1}^{N-1} C^{22}_{\frac{r}{N},1-\frac{r}{N}}\cr 
&& \quad - {1\over 2M}\sum_{r=1}^{M-1}\sum_{p>0}f^{(1)}_{\frac{r}{M},\frac{p}{M+N}}\a_{-\frac{p}{M+N}}
\sum_{q>0}f^{(1)}_{1-\frac{r}{M},\frac{q}{M+N}}\a_{-\frac{q}{M+N}}\cr
&&\quad - {1\over 2N}\sum_{r=1}^{N-1}\sum_{p>0}f^{(1)}_{\frac{r}{N},\frac{p}{M+N}}\a_{-\frac{p}{M+N}}
\sum_{q>0}f^{(1)}_{1-\frac{r}{N},\frac{q}{M+N}}\a_{-\frac{q}{M+N}}\bigg)\cr  
&&\times A\exp\Big(\sum_{p,q>0}\gamma_{\frac{p}{M+N},\frac{q}{M+N}}\a_{-\frac{p}{M+N}}\a_{-\frac{q}{M+N}}\Big)|0_{M+N}\rangle
\eea
Looking at terms without any modes, we find the relation
\bea \label{C sum}
0=e^{-w}(\partial-h)A+{1\over2M}\sum_{r=1}^{M-1} C^{11}_{\frac{r}{M},1-\frac{r}{M}}+{1\over2N}\sum_{r=1}^{N-1} C^{22}_{\frac{r}{N},1-\frac{r}{N}}  
\eea
Using (\ref{A}) we find the following constraint on the sum of contraction terms
\bea\label{C relation} 
{1\over2M}\sum_{r=1}^{M-1} C^{11}_{{r\over M},1-{r\over M}} + {1\over2N}\sum_{r=1}^{N-1} C^{22}_{{r\over N},1-{r\over N}}=e^{-w}( h - h_{M+N} + h_M + h_N )
\eea
Although we don't compute the individual values of $C^{ij}$ in this paper, we will use the above relation to simplify our computations for the quantities $\g$ and $f^{(i)}$.
Now, looking at terms with two modes $\a_{-{p\over M+N}}\a_{-{q\over M+N}}$ in (\ref{Lm1 2}), we obtain the relation
\allowdisplaybreaks
\bea
&&A\bigg( \gamma_{{p\over M+N}-1,{q\over M+N}}[L_{-1},\a_{-{p\over M+N}+1}]\a_{-{q\over M+N}}+ \gamma_{{p\over M+N},{q\over M+N}-1}\a_{-{p\over M+N }}[L_{-1},\a_{-{q\over M+N}+1}]\bigg)\nn
&&+\, \bigg(- e^{-w}(\partial - h)A \gamma_{{p\over M+N},{q\over M+N}} \nn  
&& \qquad -A \,{1\over2M}\sum_{r=1}^{M-1}f^{(1)}_{{r\over M},{p\over M+N }}f^{(1)}_{1-{r\over M},{q\over M+N }}
-A \,{1\over2N}\sum_{r=1}^{N-1}f^{(2)}_{{r\over N},{p\over M+N }}f^{(2)}_{1-{r\over N},{q\over M+N }}\nn   
&& \qquad  -A {1\over2M}\sum_{r=1}^{M-1} C^{11}_{{r\over M},1-{r\over M}}\gamma_{{p\over M+N},{q\over M+N}}-A {1\over2N}\sum_{r=1}^{N-1} C^{22}_{{r\over N},1-{r\over N}}\gamma_{{p\over M+N},{q\over M+N}}\nn
&& \qquad + A {1\over2(M+N)}\d_{\frac{p}{M+N}+\frac{q}{M+N},1}\bigg)\, \a_{-{p\over M+N}}\a_{-{q\over M+N}} = 0
\eea  
Performing the commutation relations using (\ref{L alpha multiwound}) and (\ref{C sum}) yield the relation
\bea
0\!\!&=&\!\!\big({p\over M+N}-1\big)\gamma_{{p\over M+N}-1,{q\over M+N}} + \big({q\over M+N}-1\big)\gamma_{{p\over M+N},{q\over M+N}-1}\cr  
&&- e^{-w}\partial \gamma_{{p\over M+N},{q\over M+N}}  - {1\over2M}\sum_{r=1}^{M-1}f^{(1)}_{{r\over M},{p\over M+N}}f^{(1)}_{1-{r\over M},{q\over M+N}} - {1\over2N}\sum_{r=1}^{N-1}f^{(2)}_{{r\over N},{p\over M+N}}f^{(2)}_{1-{r\over N},{q\over M+N}}\cr 
&&+ {1\over2(M+N)}\d_{{p\over M+N}+{q\over M+N},1}
\eea
Using the $w$ dependence of $\gamma$ (\ref{gam w dep M N}) we find 
\bea\label{L1 gam full}
\gamma_{{p\over M+N},{q\over M+N}}\!\!&=&\!\!e^{w}{M+N\over p+q}\bigg(\big({p\over M+N}-1\big)\gamma_{{p\over M+N}-1,{q\over M+N}}+ \big({q\over M+N}-1\big)\gamma_{{p\over M+N},{q\over M+N}-1}\cr  
&&\hspace{2.2cm} - {1\over2M}\sum_{r=1}^{M-1}f^{(1)}_{{r\over M},{p\over M+N}}f^{(1)}_{1-{r\over M},{q\over M+N}} - {1\over2N}\sum_{r=1}^{N-1}f^{(2)}_{{r\over N},{p\over M+N}}f^{(2)}_{1-{r\over N},{q\over M+N}}\cr 
&&\hspace{2.2cm} + {1\over2(M+N)}\d_{{p\over M+N}+{q\over M+N},1}\bigg)
\eea

\subsection{Relations using $L_1$}

In this section, we will derive relations using the conformal generator $L_1$. We will first consider the pair creation $\gamma$ and then the propagation $f$.

\subsubsection{Pair creation $\g$}

We start with 
\bea\label{L1 start}
0=\s(w)L_1|0_M\rangle^{(1)}|0_N\rangle^{(2)}
\eea
where we have used the fact that $L_1$ acting on the vacuum gives zero. 
Commuting $L_1$ through the twist and using (\ref{L comm}) for $n=1$ we have 
\bea \label{L1 relation}
0&=&\big(L_1\s(w) - [L_1,\s(w)]\big)|0_M\rangle^{(1)}|0_N\rangle^{(2)}\nn
&=& \big(L_1 - e^w(\partial + h)\big)\s(w)|0_M\rangle^{(1)}|0_N\rangle^{(2)}
\eea
Inserting our ansatz (\ref{chi}) gives
\be\label{L1 gamma}
0= \big(L_1 - e^w(\partial + h)\big)A\exp\Big(\sum_{p,q>0}\gamma_{{p\over M+N},{q\over M+N}}\a_{-{p\over M+N}}\a_{-{q\over M+N}}\Big)|0_{M+N}\rangle
\ee
Looking at terms without any modes, we obtain the relation
\be
0=A\sum_{p=1}^{M+N-1}\g_{{p\over M+N},1-{p\over M+N}}\big[[L_1,\a_{-{p\over M+N}}],\a_{-1+{p\over M+N}}\big] - e^w(\partial+h) A 
\ee
which gives
\be \label{gamma dimension relation}
  A \sum_{p=1}^{M+N-1}\g_{{p\over M+N},1-{p\over M+N}}\, p \, \big(1-{p\over M+N}\big)
 = e^w(\partial+h) A 
\ee
Inserting (\ref{A}) to eliminate $A$ gives
\bea \label{gamma dimension relation 1}
  \sum_{p=1}^{M+N-1}\g_{{p\over M+N},1-{p\over M+N}}\, p \, \big(1-{p\over M+N}\big)
 = e^w(h+h_{M+N}-h_M-h_N)
\eea
Next, looking at terms with two modes $\a_{-{p\over M+N}}\a_{-{q\over M+N}}$ in (\ref{L1 gamma}), we obtain
\bea 
0&=&A\g_{1 + {p\over M+N},{q\over M+N}} [L_1,\a_{-1-{p\over M+N}}]\a_{-{q\over M+N}}+ A\g_{{p\over M+N},1+{q\over M+N}}\a_{-{p\over M+N}}[L_1,\a_{-1-{q\over M+N}}] \nn
&& + A\sum_{p'=1}^{M+N-1}\g_{{p\over M+N},{q\over M+N}}\g_{{p'\over M+N},1-{p'\over M+N}}\big[[L_1,\a_{-{p'\over M+N}}],\a_{-1+{p'\over M+N}}\big]\a_{-{p\over M+N}}\a_{-{q\over M+N}}
\nn 
&& + 2A\sum_{p'=1}^{M+N-1}\g_{{p\over M+N},{p'\over M+N}}\g_{{q\over M+N},1-{p'\over M+N}}\big[[L_1,\a_{-{p'\over M+N}}],\a_{-1+{p'\over M+N}}\big]\a_{-{p\over M+N}}\a_{-{q\over M+N}}\nn
&&- e^w(\partial+h) A\g_{{p\over M+N},{q\over M+N}}\a_{-{p\over M+N}}\a_{-{q\over M+N}}
\eea
Performing the commutation relations (\ref{L alpha multiwound}) yield
\bea
0&=&A\g_{1 + {p\over M+N},{q\over M+N}}\big(1+{p\over M+N}\big)+ A\g_{{p\over M+N},1+{q\over M+N}}(1+{q\over M+N}) \nn  
&& + A\sum_{p'=1}^{M+N-1}\g_{{p\over M+N},{q\over M+N}}\g_{{p'\over M+N},1-{p'\over M+N}}\,p'\,\big(1-{p'\over M+N}\big)\cr  
&& + 2A\sum_{p'=1}^{M+N-1}\g_{{p\over M+N},{p'\over M+N}}\g_{{q\over M+N},1-{p'\over M+N}}\,p'\,\big(1-{p'\over M+N}\big)\nn 
&&- e^w(\partial+h) A\g_{{p\over M+N},{q\over M+N}}
\eea
Using (\ref{gamma dimension relation}) to eliminate $A$, we get the relation
\bea 
e^w\partial\g_{{p\over M+N},{q\over M+N}}\!\!&=&\!\! \g_{1 + {p\over M+N},{q\over M+N}}\big(1+{p\over M+N}\big)  + \g_{{p\over M+N},1+{q\over M+N}}\big(1+{q\over M+N}\big)\cr
&&+2\sum_{p'=1}^{M+N-1}\g_{{p\over M+N},{p'\over M+N}}\g_{{q\over M+N},1-{p'\over M+N}}\, p'\, \big(1-{p'\over M+N}\big)
\eea
Inserting the $w$ dependence of $\gamma$ given in (\ref{gam w dep M N}) we obtain the relation
\bea\label{L1 gam} 
\g_{{p\over M+N},{q\over M+N}}
\!\!&=&\!\! e^{-w}\,{M+N\over p+q}\bigg(\g_{1 + {p\over M+N},{q\over M+N}}\big(1+{p\over M+N}\big)  + \g_{{p\over M+N},1+{q\over M+N}}\big(1+{q\over M+N}\big)\cr
&&\hspace{2.2cm}+\,2\sum_{p'=1}^{M+N-1}\g_{{p\over M+N},{p'\over M+N}}\g_{{q\over M+N},1-{p'\over M+N}}\, p'\, \big(1-{p'\over M+N}\big)\bigg)
\eea

\subsubsection{Propagation $f$}

Now let us derive relations for the propagation $f$ using $L_1$. We first compute the relations for $f^{(1)}$ and then note that the relations for $f^{(2)}$ can be obtained in a similar way. We start with the following equation
\bea 
0 \!&=&\!  \s(w)L_1\a_{-{r\over M}}|0_M\rangle^{(1)}|0_N\rangle^{(2)},\qquad \qquad\qquad 1\leq r\leq M-1\cr  
\!&=&\!  \big(L_1 - e^w(\partial + h)\big) \s(w)\a_{-{r\over M}}|0_M\rangle^{(1)}|0_N\rangle^{(2)}\cr
\!&=&\!  \big(L_1 - e^w(\partial + h)\big) \sum_{p'> 0}f^{(1)}_{{r\over M},{p'\over M+N}}\a_{-{p'\over M+N}} A\exp\Big(\sum_{p,q>0}\gamma_{{p\over M+N},{q\over M+N}}\a_{-{p\over M+N}}\a_{-{q\over M+N}}\Big)|0_{M+N}\rangle \nn
\eea
Looking at terms with a single mode $\a_{-{p\over M+N}}$, we have
\allowdisplaybreaks
\bea 
0&=&A\,f^{(1)}_{{r\over M},1+{p\over M+N}}[L_1 ,\a_{-1-{p\over M+N}} ]\nn  
&& +\  A \sum_{q=1}^{M+N-1}\big(2f^{(1)}_{{r\over M},{q\over M+N}}\gamma_{1-{q\over M+N},{p\over M+N}}+f^{(1)}_{{r\over M},{p\over M+N}}\gamma_{{q\over M+N},1-{q\over M+N}}\big)\nn
&& \qquad \qquad \ \times\,\big[[L_1,\a_{-{q\over M+N}}],\a_{-1+{q\over M+N}}\big]\a_{-{p\over M+N}}\nn  
&&-\ e^w(\partial + h)A\,f^{(1)}_{{r\over M},{p\over M+N}}\a_{-{p\over M+N}}
\eea
Performing commutation relations (\ref{L alpha multiwound}) give
\bea 
0&=&A\, f^{(1)}_{{r\over M},1+{p\over M+N}}\big(1+{p\over M+N}\big)\cr  
&& +\  A \sum_{q=1}^{M+N-1}\big(2f^{(1)}_{{r\over M},{q\over M+N}}\gamma_{1-{q\over M+N},{p\over M+N}}+f^{(1)}_{{r\over M},{p\over M+N}}\gamma_{{q\over M+N},1-{q\over M+N}}\big)\,q\,\big(1-{q\over M+N}\big)\nn
&& -\ e^w(\partial + h)A\,f^{(1)}_{{r\over M},{p\over M+N}}
\eea
Inserting the expression (\ref{gamma dimension relation}) to eliminate $A$ gives
\bea 
&& f^{(1)}_{{r\over M},1+{p\over M+N}}\big(1+{p\over M+N}\big)  \nn
 &=& e^w\partial f^{(1)}_{{r\over M},{p\over M+N}} - \sum_{q=1}^{M+N-1}2\,f^{(1)}_{{r\over M},{q\over M+N}}\gamma_{1-{q\over M+N},{p\over M+N}}\,q\,\big(1-{q\over M+N}\big)
\eea
Using the relation (\ref{f dep w}), which provides the $w$-dependence of $f$, we find the relation
\bea \label{f1}
&&\hspace{-0.5cm}f^{(1)}_{{r\over M},1+{p\over M+N}}\big(1+{p\over M+N}\big) \nn
 &&\hspace{-1cm}=\,  e^w\big({p\over M+N} - {r\over M} \big) f^{(1)}_{{r\over M},{p\over M+N}} - 2\sum_{q=1}^{M+N-1}f^{(1)}_{{r\over M},{q\over M+N}}\gamma_{1-{q\over M+N},{p\over M+N}}\,q\,\big(1-{q\over M+N}\big) \
\eea
Similarly, $f^{(2)}$ can be obtained by taking $M \leftrightarrow N$
\bea \label{f2}
&&\hspace{-0.5cm}f^{(2)}_{{r\over N},1+{p\over M+N}}\big(1+{p\over M+N}\big) \nn
 &&\hspace{-1cm}=\, e^w\big({p\over M+N} - {r\over N} \big) f^{(2)}_{{r\over N},{p\over M+N}} - 2\sum_{q=1}^{M+N-1}f^{(2)}_{{r\over N},{q\over M+N}}\gamma_{1-{q\over M+N},{p\over M+N}}\,q\,\big(1-{q\over M+N}\big)
\eea

\subsection{Relations using $L_{n>0}$}

Here we use the generator, $L_n$, for $n>0$, to obtain additional relations for $\gamma$. We start with the following relation
\bea 
0=\s(w)L_{n}|0_{M}\rangle^{(1)}|0_{N}\rangle^{(2)} , \qquad n>0
\eea
Acting $L_n$ to the left and using (\ref{L comm}) we find
\bea
\label{Ln}
0&=&\big(L_{n}\,\s(w)-[L_{n},\s(w)]\big)|0_{M}\rangle^{(1)}|0_{N}\rangle^{(2)}  \nn
&=&\big(L_n -  e^{nw}(\partial+nh)\big)A\exp\Big(\sum_{p,q>0}\gamma_{{p\over M+N},{q\over M+N}}\a_{-{p\over M+N}}\a_{-{q\over M+N}}\Big)|0_{M+N}\rangle
\eea

Looking at terms without any modes, we find the relation 
\bea
0\!\!&=&\!\! A\!\sum_{p=1}^{n(M+N)-1}\!\gamma_{{p\over M+N},n-{p\over M+N}}\big[[L_n,\a_{-{p\over M+N}}],\a_{-n+{p\over M+N}}\big]-e^{nw}(\partial +nh)A 
\cr  
\!\!&=&\!\! A\!\sum_{p=1}^{n(M+N)-1}\!\gamma_{{p\over M+N},n-{p\over M+N}}\, p\, \big(n-{p\over M+N}\big)
-e^{nw}(nh + h_{M+N}-h_M-h_N)A 
\eea 
which gives the relation
\bea \label{Ln gam}
&&\sum_{p=1}^{n(M+N)-1}\gamma_{{p\over M+N},n-{p\over M+N}}\,p\, \big(n-{p\over M+N}\big) = e^{nw}(nh + h_{M+N}-h_M-h_N)
\eea
We note that for $n=1$, relation (\ref{Ln gam}) reduces to (\ref{gamma dimension relation 1}).
So far we have used the Virasoro generators, $L_{-1},L_0,L_1,L_{n>0}$ to find a set of relations that need to be solved. We summarize these relations in the next section.

\section{Summary of Results}\label{section 6}

In this section, we collect all the relevant equations and classify them into two categories: recursion relations and constraints. First notice that the $w$-dependence is given by the relations derived from the generator $L_0$ (\ref{gam w dep M N}), (\ref{f dep w}), and (\ref{f dep w 2})
\bea\label{w depend}
\gamma_{{p\over M+N},{q\over M+N}}&\propto&e^{({p+q\over M+N})w}\nn
f^{(1)}_{{r\over M},{p\over M+N}}&\propto& e^{({p\over M+N}-{r\over M})w}\nn
f^{(2)}_{{r\over N},{p\over M+N}}&\propto& e^{({p\over M+N}-{r\over N})w}
\eea

\subsection{Recursion relations}\label{sec recursion}

The following relations are categorized as recursion relations: the relations derived from the generator $L_{-1}$ from terms with two modes (\ref{L1 gam full})
\bea\label{r1}
\gamma_{{p\over M+N},{q\over M+N}}\!\!&=&\!\!e^{w}{M+N\over p+q}\bigg(\big({p\over M+N}-1\big)\gamma_{{p\over M+N}-1,{q\over M+N}}+ \big({q\over M+N}-1\big)\gamma_{{p\over M+N},{q\over M+N}-1}\cr  
&&\hspace{2.2cm} - {1\over2M}\sum_{r=1}^{M-1}f^{(1)}_{{r\over M},{p\over M+N}}f^{(1)}_{1-{r\over M},{q\over M+N}} - {1\over2N}\sum_{r=1}^{N-1}f^{(2)}_{{r\over N},{p\over M+N}}f^{(2)}_{1-{r\over N},{q\over M+N}}\cr 
&&\hspace{2.2cm} + {1\over2(M+N)}\d_{{p\over M+N}+{q\over M+N},1}\bigg)
\eea
and the relations derived from the generator $L_{1}$ for propagation $f^{(1)}$ (\ref{f1}) and $f^{(2)}$ (\ref{f2})
\bea \label{r2}
&&\hspace{-0.5cm}f^{(1)}_{{r\over M},1+{p\over M+N}}\big(1+{p\over M+N}\big) \nn
 &&\hspace{-1cm}=\,  e^w\big({p\over M+N} - {r\over M} \big) f^{(1)}_{{r\over M},{p\over M+N}} - 2\sum_{q=1}^{M+N-1}f^{(1)}_{{r\over M},{q\over M+N}}\gamma_{1-{q\over M+N},{p\over M+N}}\,q\,\big(1-{q\over M+N}\big) \
\eea
and
\bea \label{r3}
&&\hspace{-0.5cm}f^{(2)}_{{r\over N},1+{p\over M+N}}\big(1+{p\over M+N}\big) \nn
 &&\hspace{-1cm}=\, e^w\big({p\over M+N} - {r\over N} \big) f^{(2)}_{{r\over N},{p\over M+N}} - 2\sum_{q=1}^{M+N-1}f^{(2)}_{{r\over N},{q\over M+N}}\gamma_{1-{q\over M+N},{p\over M+N}}\,q\,\big(1-{q\over M+N}\big)
\eea
Given all the values of $f^{(1)}_{\frac{r}{M}<1,\frac{p}{M+N}<1}$ and $f^{(2)}_{\frac{r}{N}<1,\frac{p}{M+N}<1}$, we can determine the values of $\gamma_{\frac{p}{M+N},\frac{q}{M+N}}$, $f^{(1)}_{\frac{r}{M}<1,\frac{p}{M+N}}$, and $f^{(2)}_{\frac{r}{N}<1,\frac{p}{M+N}}$ by the following steps:

\b

1) Use relations (\ref{r1}) to determine all $\gamma_{\frac{p}{M+N},\frac{q}{M+N}}$ where $\frac{p}{M+N},\frac{q}{M+N}< 1$. 

\b

2) Use relations (\ref{r2}) and (\ref{r3}) to determine all $f^{(1)}_{\frac{r}{M}<1,\, 1<\frac{p}{M+N}<2}$
and $f^{(2)}_{\frac{r}{N}<1,\, 1<\frac{p}{M+N}<2}$.  

\b

3) Use relations (\ref{r1}) to determine all  $\gamma_{\frac{p}{M+N},\frac{q}{M+N}}$ where $\frac{p}{M+N},\frac{q}{M+N}< 2$. 

\b

4) Use relations (\ref{r2}) and (\ref{r3}) to determine all $f^{(1)}_{\frac{r}{M}<1,\, 2<\frac{p}{M+N}<3}$
and $f^{(2)}_{\frac{r}{N}<1,\, 2<\frac{p}{M+N}<3}$.

\b

$\dots$

\b 

Therefore, using these recursion relations, the infinite number of coefficients for the pair creation can be determined by a finite number of inputs
\be\label{inputs}
f^{(1)}_{\frac{r}{M}<1,\frac{p}{M+N}<1}\, , \ f^{(2)}_{\frac{r}{N}<1,\frac{p}{M+N}<1}\  \implies \ \gamma_{\frac{p}{M+N},\frac{q}{M+N}}
\ee
with the number given by 
\be\label{N 0 inputs}
(M-1)(M+N-1)+(N-1)(M+N-1)= (M+N-2)(M+N-1)
\ee
Next, we write down constraint equations that can be used to reduce the number of inputs.

\subsection{Constraints}\label{subsection constraints}

There are also constraints that help us to reduce the number of required inputs or even determine them. The first type of constraints are given by the relations derived from the generator $L_{n>0}$ for $\gamma$ (\ref{Ln gam}), which comes from terms without any modes
\bea \label{c1}
&&\sum_{p=1}^{n(M+N)-1}\gamma_{{p\over M+N},n-{p\over M+N}}\,p\, \big(n-{p\over M+N}\big) = e^{nw}(nh + h_{M+N}-h_M-h_N)
\eea
The second type of constraints are given by (\ref{L1 gam}), which comes from the generator $L_{1}$ for $\gamma$ from terms with two modes
\bea\label{c2} 
\g_{{p\over M+N},{q\over M+N}}
\!\!&=&\!\! e^{-w}\,{M+N\over p+q}\bigg(\g_{1 + {p\over M+N},{q\over M+N}}\big(1+{p\over M+N}\big)  + \g_{{p\over M+N},1+{q\over M+N}}\big(1+{q\over M+N}\big)\cr
&&\hspace{2.2cm}+\,2\sum_{p'=1}^{M+N-1}\g_{{p\over M+N},{p'\over M+N}}\g_{{q\over M+N},1-{p'\over M+N}}\, p'\, \big(1-{p'\over M+N}\big)\bigg)
\eea
Unlike the recursion relations that determine higher energy $\gamma$ and $f$ in terms of lower energy quantities, these constraints provide relations between $\gamma$ at the same energy, expressed in terms of $\gamma$ at lower energy. For example, (\ref{c2}) gives constraints for $\g_{1 + {p\over M+N},{q\over M+N}}$ and $\g_{{p\over M+N},1+{q\over M+N}}$ in terms of the lower energy $\g_{ {p\over M+N},{q\over M+N}}$ and $\sum_{p'=1}^{M+N-1}\g_{{p\over M+N},{p'\over M+N}}\g_{{q\over M+N},1-{p'\over M+N}}$.

Notice that as we include higher energy constraints (e.g., larger $n$ in (\ref{c1}) and larger $p$ and $q$ in (\ref{c2})), the number of constraints will increase. At a critical energy bound, naively, it would seem that we would have more constraints than inputs (\ref{inputs}) providing the possibility of determining all required inputs without having to insert any at the beginning. 
It is clear that not all constraints are independent. 
As we will show with explicit examples, it seems that the constraints coming from (\ref{c2}) are all trivial and thus do not help in solving for the relevant coefficients. They only serve as an additional check that the correct equations are being used. It appears that constraints (\ref{c1})  are nontrivial. Using the constraints, the number of inputs (\ref{N 0 inputs}) can be reduced to
\begin{align}\label{N inputs}
N^{\text{inputs}}=(M+N-2)(M+N-1) - N^{\text{constr}}
\end{align}
In section \ref{section 8}, we will provide explicit examples for $(M,N)=(2,1), (3,1), (2,2)$. We will show that indeed some of the constraints are nontrivial and effectively reduce the number of required inputs.

\section{Results from Covering Map}\label{section 7}

Here we record relevant results for pair creation $\gamma$ and propagation $f$, which were obtained using the covering map method \cite{Avery:2010er,Avery:2010hs,Avery:2010qw,Carson:2014yxa,Carson:2014ena}. 
In \cite{Carson:2014yxa}, the coefficients $\gamma$ and $f$ were computed in the D1D5 CFT with four free bosons and four free fermions, considering the same twist configuration as studied in this paper: a single $\s(w)$ twists together an initial copy of winding $M$ and an initial copy of winding $N$ into a copy of winding $M+N$.
In our case, although we only consider a single real boson, we can use the results from \cite{Carson:2014yxa} because all the bosons are free in both cases. However, an appropriate rescaling factor is needed since the bosons used in \cite{Carson:2014yxa} are complex bosons that consist of two real bosons each. Therefore, we have
\bea \label{MN gamma}
&&\gamma_{{p\over M+N},{q\over M+N}}={1\over2}\gamma^{\text{D1D5}}_{{p\over M+N},{q\over M+N}}\nn
&=&-{1\over2}{a^{p+q}\over\pi^2}\sin\big[{\pi N p\over M+N}\big]\sin\big[{\pi N q\over M+N}\big]{MN\over (M+N)^2}{1\over p + q}{\Gamma[{Mp\over M+N}]\Gamma[{Np\over M+N}]\over\Gamma[p]}{\Gamma[{Mq\over M+N}]\Gamma[{Nq\over M+N}]\over\Gamma[q]}\nn
\eea
As for the $f^{(i)}$, it remains unchanged since it computes the transition from one boson in the initial state to one boson in the final state. The rescaling factors cancel out and there is no overall rescaling factor left. The expressions are given by 
\bea\label{f1}
f^{(1)}_{{r\over M},{p\over M+N}} &=& f^{(1),\text{D1D5}}_{{r\over M},{p\over M+N}}\nn 
&=&{(-1)^r\sin({\pi M p\over M+N})\over\pi(M+N)}{(-a)^{p-{(M+N)r\over M}}\over {p\over M+N}-{r\over M}}{\Gamma[{(M+N)r\over M}]\over\Gamma[r]\Gamma[{Nr\over M}]}
{\Gamma[{Mp\over M+N}]\Gamma[{Np\over M+N}]\over\Gamma[p]},\quad{r\over M} \neq {p\over M+N}
\nn
\nn
f^{(1)}_{{r\over M},{p\over M+N}} &=& {M\over M+N},\quad  {r\over M} = {p\over M+N}
\eea
and
\bea\label{f2}
f^{(2)}_{{r\over N},{p\over M+N}} &=& f^{(2),\text{D1D5}}_{{r\over N},{p\over M+N}}\nn 
&=&{(-1)^r\sin({\pi N p\over M+N})\over\pi(M+N)}{(-a)^{p-{(M+N)r\over N}}\over {p\over M+N}-{r\over N}}{\Gamma[{(M+N)r\over N}]\over\Gamma[r]\Gamma[{Mr\over N}]}
{\Gamma[{Mp\over M+N}]\Gamma[{Np\over M+N}]\over\Gamma[p]},\quad{r\over N} \neq {p\over M+N}
\nn
\nn
f^{(2)}_{{r\over N},{p\over M+N}} &=& {N\over M+N},\quad  {r\over N} = {p\over M+N}
\eea
where
\be
a=e^{-i\pi{N\over M+N}}\bigg({e^w\over M^M N^N}\bigg)^{1\over M+N}(M+N) 
\ee
Here, $w$ is the location of $\s(w)$.
In the next section, we provide several examples where we solve the recursion relations and constraints derived in the previous sections for specific values of initial winding $M$ and $N$. 
We will show that the resulting pair creation $\gamma$ and propagation $f$ are the same as the results in this section.

\section{Examples}\label{section 8}

In this section, we provide examples of solving the recursion relations and constraints for $(M,N)=(2,1), (3,1), (2,2)$. Notice that the constraints (\ref{c1}) and (\ref{c2}) include only the pair creation $\gamma$. However, it remains unclear which constraints are trivial and which are nontrivial. To explore this, we consider some explicit examples and restrict ourselves to constraints that contain only the following $\gamma$
\bea 
\gamma_{{p\over M+N}<2,{q\over M+N}<2},\qquad {p\over M+N}+{q\over M+N}\leq 2
\eea
Therefore we only need to consider the constraint (\ref{c1}) for $n=1,2$, as well as the lowest constraints in (\ref{c2}) with $\frac{p}{M+N}+\frac{q}{M+N}\leq 1$. To determine these $\gamma$ from the recursion relations, we will also need to find 
\be
f^{(i)}_{{r\over M}<1,{p\over M+N}<2},\quad i=1,2
\ee
Notice that since we are considering only some of the lowest constraints, not all powers of the constraints have been utilized. Nevertheless, by explicitly solving these examples, we can gain some insight into which constraints are nontrivial, leading us to a better understanding of the underlying structures. 

In the case of $(M,N)=(2,1)$, we find that there is only one nontrivial constraint $N^{\text{constr}}=1$ coming from the first type (\ref{c1}) with $n=1$. For $(M,N)=(3,1),(2,2)$, there are two nontrivial constraints $N^{\text{constr}}=2$ coming from the first type (\ref{c1}) with $n=1,2$.

\subsection{$M=2,N=1$}

In this case, the $w$-dependence is given by
\bea\label{case 1 w}
\gamma_{{p\over3},{q\over3}}&\propto& e^{{p+q\over 3}w}\cr  
f^{(1)}_{{r\over2},{p\over3}}&\propto& e^{({p\over3} - {r\over2})w}\cr 
f^{(2)}_{r,{p\over3}}&\propto&e^{({p\over3} - r)w}
\eea
The recursion relations (\ref{r1}) and (\ref{r2}) are given by
\be\label{gam greater one 2 1}
\gamma_{{p\over3},{q\over3}}=e^w{3\over p+q}\bigg(\big({p\over3}-1\big)\gamma_{{p\over3}-1,{q\over3}} + \big({q\over3}-1\big)\gamma_{{p\over3},{q\over3}-1} - {1\over4}f^{(1)}_{{1\over2},{p\over3}}f^{(1)}_{{1\over2},{q\over3}} +\frac{1}{6}\delta_{\frac{p}{3}+\frac{q}{3},1}\bigg)
\ee
and
\be \label{f1 2 1}
f^{(1)}_{{r\over2},1+{p\over3}}\big(1+{p\over3}\big)=e^w\big({p\over3} - {r\over2} \big)f^{(1)}_{{r\over2},{p\over3}}- {4\over3}f^{(1)}_{{r\over2},{1\over3}}\gamma_{{2\over3},{p\over3}}  - {4\over3}f^{(1)}_{{r\over2},{2\over3}}\gamma_{{1\over3},{p\over3}}
\ee
We note that for $N=1$ we do not get any $f^{(2)}$ terms.

The constraints (\ref{c1}) with $n=1,2$ and (\ref{c2}) become
\be \label{gam rel}
\gamma_{{1\over 3},{2\over 3}}={1\over12}e^w
\ee
\be \label{L2 M2N1}
\sum_{p=1}^{5}\gamma_{{p\over3},2-{p\over3}}\, p\, (2-{p\over3}) = {25\over 144}e^{2w}
\ee
\be \label{L1 gam 2 1}
\g_{{p\over3},{q\over3}} = e^{-w}{3\over p+q}\bigg(\g_{1 + {p\over 3},{q\over 3}}\big(1+{p\over 3}\big)  + \g_{{p\over 3},1+{q\over 3}}\big(1+{q\over 3}\big)+{4\over3}\g_{{p\over 3},{1\over 3}}\g_{{2\over 3},{q\over 3}} + {4\over3}\g_{{p\over 3},{2\over 3}}\g_{{1\over 3},{q\over 3}}\,\bigg)
\ee

\subsubsection*{Explicit relations}\label{subsubsection M2N1}

The set of coefficients we consider is 
\bea 
\gamma_{{m\over3},{n\over3}},\quad \ {m\over3}+{n\over3} \leq 2 \qquad\text{ and }\qquad f^{(1)}_{{r\over2},{m\over3}}, \quad \ {r\over2} < 1,\ {m\over3} < 2
\eea
which are the following 10 variables
\bea
&&\gamma_{{1\over3},{1\over3}},\ \gamma_{{1\over3},{2\over3}},\ \gamma_{{2\over3},{2\over3}},\ \gamma_{{1\over3},{4\over3}},\ \gamma_{{1\over3},{5\over3}},\ \gamma_{{2\over3},{4\over3}},\ f^{(1)}_{{1\over2},{1\over3}},\ f^{(1)}_{{1\over2},{2\over3}},\ f^{(1)}_{{1\over2},{4\over3}},\ f^{(1)}_{{1\over2},{5\over3}}
\eea
At each step in section \ref{sec recursion}, we keep equations containing only the above variables. It is sufficient to proceed up to step 3) in the recursion relations. In the following analysis, we will set $w=0$, and we can reintroduce the $w$-dependence using (\ref{case 1 w}). The recursion relations are 
\begin{align}\label{M2N1 eqn 1}
\text{step 1)}\qquad\qquad\gamma _{\frac{1}{3},\frac{1}{3}}&=-\frac{3}{8} (f^{(1)}_{\frac{1}{2},\frac{1}{3}})^2  \\[6pt]
\gamma_{\frac{1}{3},\frac{2}{3}}&=\frac{1}{6}-\frac{1}{4} f^{(1)}_{\frac{1}{2},\frac{1}{3}} f^{(1)}_{\frac{1}{2},\frac{2}{3}} \\[6pt]
\label{M2N1 eqn 3}
\gamma _{\frac{2}{3},\frac{2}{3}}&=-\frac{3}{16} (f^{(1)}_{\frac{1}{2},\frac{2}{3}})^2 \\[6pt]
\label{M2N1 eqn 4}
\text{step 2)}\qquad\quad\ \frac{4}{3} f^{(1)}_{\frac{1}{2},\frac{4}{3}}&=-\frac{1}{6} f^{(1)}_{\frac{1}{2},\frac{1}{3}}- \frac{4}{3} \gamma _{\frac{1}{3},\frac{1}{3}}f^{(1)}_{\frac{1}{2},\frac{2}{3}}- \frac{4}{3} \gamma_{\frac{1}{3},\frac{2}{3}} f^{(1)}_{\frac{1}{2},\frac{1}{3}}  \\[6pt]
\label{M2N1 eqn 5}
\frac{5}{3}f^{(1)}_{\frac{1}{2},\frac{5}{3}}&=\ \frac{1}{6} f^{(1)}_{\frac{1}{2},\frac{2}{3}}- \frac{4}{3} \gamma _{\frac{1}{3},\frac{2}{3}}
f^{(1)}_{\frac{1}{2},\frac{2}{3}}-\frac{4}{3} \gamma _{\frac{2}{3},\frac{2}{3}} f^{(1)}_{\frac{1}{2},\frac{1}{3}}  \\[6pt]
\label{M2N1 eqn 6}
\text{step 3)}\qquad\qquad\gamma_{\frac{1}{3},\frac{4}{3}}&=\frac{3}{5} \left(\frac{1}{3} \gamma _{\frac{1}{3},\frac{1}{3}}-\frac{1}{4} f^{(1)}_{\frac{1}{2},\frac{1}{3}}f^{(1)}_{\frac{1}{2},\frac{4}{3}}\right) \\[6pt]
\gamma _{\frac{1}{3},\frac{5}{3}}&=\frac{1}{2} \left(\frac{2}{3} \gamma
_{\frac{1}{3},\frac{2}{3}}-\frac{1}{4} f^{(1)}_{\frac{1}{2},\frac{1}{3}} f^{(1)}_{\frac{1}{2},\frac{5}{3}}\right) \\[6pt]\label{M2N1 eqn 8}
\gamma_{\frac{2}{3},\frac{4}{3}}&=\frac{1}{2} \left(\frac{1}{3} \gamma _{\frac{1}{3},\frac{2}{3}}-\frac{1}{4} f^{(1)}_{\frac{1}{2},\frac{2}{3}}f^{(1)}_{\frac{1}{2},\frac{4}{3}}\right)
\end{align}
The constraints are
\begin{align}\label{constraint 1}
\frac{1}{9} &= \frac{4}{3} \gamma _{\frac{1}{3},\frac{2}{3}}  \\[6pt]
\label{constraint 4}
\frac{25}{144}&=\frac{16}{3} \gamma _{\frac{2}{3},\frac{4}{3}}+\frac{10}{3} \gamma_{\frac{1}{3},\frac{5}{3}}\\[6pt]
\label{constraint 2}
\gamma_{\frac{1}{3},\frac{1}{3}}&=\frac{3}{2} \left(\frac{8}{3} \gamma _{\frac{1}{3},\frac{1}{3}} \gamma_{\frac{1}{3},\frac{2}{3}}+\frac{8}{3} \gamma _{\frac{1}{3},\frac{4}{3}}\right)  \\[6pt]
\label{constraint 3}
\gamma _{\frac{1}{3},\frac{2}{3}}&=\frac{4}{3} \gamma _{\frac{1}{3},\frac{2}{3}}^2+\frac{4}{3} \gamma _{\frac{1}{3},\frac{1}{3}} \gamma_{\frac{2}{3},\frac{2}{3}}+\frac{4}{3} \gamma _{\frac{2}{3},\frac{4}{3}}+\frac{5}{3} \gamma_{\frac{1}{3},\frac{5}{3}}  
\end{align}

By using the recursion relations, all 10 variables can be determined using just 2 variables, as indicated in (\ref{inputs})
\bea
f^{(1)}_{{1\over2},{1\over3}},\ f^{(1)}_{{1\over2},{2\over3}}
\eea
We have 4 constraints to determine them. To show this more clearly, we first make the following definitions
\bea\label{definitions}
f^{(1)}_{{1\over2},{1\over3}}\equiv x,\qquad
 f^{(1)}_{{1\over2},{2\over3}}\equiv y
\eea

\subsubsection*{Step $1$}

Inserting these definitions into the equations involved in step $1$ (\ref{M2N1 eqn 1} -- \ref{M2N1 eqn 3}) 
yield 
\begin{align} \label{step 1 a}
\gamma _{\frac{1}{3},\frac{1}{3}}&= -\frac{3 x^2}{8}\\
\label{step 1 b}
\gamma _{\frac{1}{3},\frac{2}{3}}&=\frac{1}{6}-\frac{x y}{4}\\ 
\label{step 1 c}
\gamma_{\frac{2}{3},\frac{2}{3}}&=-\frac{3 y^2}{16}
\end{align}
\subsubsection*{Step $2$}
Inserting the definitions (\ref{definitions}) and the relations (\ref{step 1 a} -- \ref{step 1 c}) into the relations for step $2$ (\ref{M2N1 eqn 4}), (\ref{M2N1 eqn 5}), give
\begin{align} \label{step 2 a}
f_{\frac{1}{2},\frac{4}{3}}&= \frac{1}{24} x (15 x y-7)\\ 
\label{step 2 b}
f_{\frac{1}{2},\frac{5}{3}}&= \frac{1}{60} y(21 x y-2)
\end{align}
\subsubsection*{Step $3$}
Inserting definitions (\ref{definitions}) and the relations from step 1 (\ref{step 1 a} -- \ref{step 1 c}) and step 2 (\ref{step 2 a}), (\ref{step 2 b}) into the relations for step 3 (\ref{M2N1 eqn 6} -- \ref{M2N1 eqn 8}) yield
\begin{align}\label{step 3 a}
\gamma _{\frac{1}{3},\frac{4}{3}}&= -\frac{1}{32} x^2 (3 x y+1)
\\
\gamma_{\frac{1}{3},\frac{5}{3}}&= \frac{1}{18}-\frac{1}{480} x y (21 x y+38)
\\
\label{step 3 c}
\gamma _{\frac{2}{3},\frac{4}{3}}&= \frac{1}{576} \big(16-3 x y (15 xy+1)\big)
\end{align}

\subsubsection*{Constraints}

Inserting the expressions (\ref{step 1 a} -- \ref{step 3 c}) into the constraint equations (\ref{constraint 1} - \ref{constraint 4}) 
we find that it gives one nontrivial relation coming from (\ref{constraint 1}) 
\begin{align} \label{real constraint}
1=3xy
\end{align}
The other constraints turn out to be trivial.  

\subsubsection*{Solutions}

Inserting the constraint (\ref{real constraint}) into (\ref{step 1 a} -- \ref{step 3 c}) we obtain all 10 variables in terms of a single variable, $x=f^{(1)}_{{1\over2},{1\over3}}$,
\begin{align} \label{relation}
\gamma _{\frac{1}{3},\frac{2}{3}}&=\frac{1}{12},&\gamma _{\frac{1}{3},\frac{5}{3}}&=\frac{7}{288},&\gamma _{\frac{2}{3},\frac{4}{3}}&=\frac{5}{288}\nn
\nn
\gamma_{\frac{1}{3},\frac{1}{3}}&=-\frac{3 (f^{(1)}_{{1\over2},{1\over3}})^2}{8},&\gamma _{\frac{1}{3},\frac{4}{3}}&= -\frac{(f^{(1)}_{{1\over2},{1\over3}})^2}{16},&\gamma_{\frac{2}{3},\frac{2}{3}}&= -\frac{1}{48 (f^{(1)}_{{1\over2},{1\over3}})^2},\nn  
f_{\frac{1}{2},\frac{2}{3}}&={1\over3f^{(1)}_{{1\over2},{1\over3}}},&f_{\frac{1}{2},\frac{4}{3}}&= -\frac{f^{(1)}_{{1\over2},{1\over3}}}{12},&f_{\frac{1}{2},\frac{5}{3}}&= \frac{1}{36f^{(1)}_{{1\over2},{1\over3}}}
\end{align}
We notice, written in the first row, that the $\gamma$'s whose indices obey the relation ${p\over 3}+{q\over 3}\in\mathbb{Z}$ are completely solved without the need for extra input. 
These values are in agreement with (\ref{MN gamma}). In order to solve for the remaining $\gamma$'s and $f$'s we need a single input which, using (\ref{f1}), we write as  
\be\label{input} 
f^{(1)}_{{1\over2},{1\over3}}=-{(-4)^{2\over3}\over\sqrt{3}}
\ee
Inserting this into the second and third row of (\ref{relation}) the remaining values of $\gamma$ and $f$ are given by
\begin{align} 
\gamma_{\frac{1}{3},\frac{1}{3}}&=\frac{({-\frac{1}{2}})^{1\over3}}{4} ,&\gamma_{\frac{1}{3},\frac{4}{3}}&= \frac{({-\frac{1}{2}})^{1\over3}}{24},&\gamma _{\frac{2}{3},\frac{2}{3}}&=-\frac{(-\frac{1}{2})^{2\over3}}{16}\nn 
f_{\frac{1}{2},\frac{2}{3}}&={(-{1\over2})^{1\over3}\over\sqrt3},&f_{\frac{1}{2},\frac{4}{3}}&=\frac{(-{1\over2})^{2\over3}}{6~\sqrt{3}},&f_{\frac{1}{2},\frac{5}{3}}&= \frac{({-\frac{1}{2}})^{1\over3}}{12 \sqrt{3}}
\end{align}
which also agree with (\ref{MN gamma}) and (\ref{f1}). This shows that the equations derived from conformal generators are indeed correct. We note that the number of inputs is 1, in agreement with equation (\ref{N inputs}) for $(M,N)=(2,1)$ and $N^{\text{constr}}=1$.  
Next, we look at the case $(M,N)=(3,1)$ with low energy constraints.

\subsection{$M=3,N=1$}

In this case, the $w$-dependence is given by
\bea\label{case 2 w}
\gamma_{{p\over4},{q\over4}}&\propto& e^{{p+q\over4} w}\cr  
f^{(1)}_{{r\over3},{p\over4}}&\propto& e^{({p\over4} - {r\over3})w}\cr 
f^{(2)}_{r,{p\over4}}&\propto&e^{({p\over4} - r)w}
\eea
The recursion relations (\ref{r1}), (\ref{r2}), and (\ref{r3}) are given by
\begin{align}\label{gam greater one 2 1 M3}
\gamma_{{p\over4},{q\over4}}&=e^w{4\over p+q}\bigg(\big({p\over4}-1\big)\gamma_{{p\over4}-1,{q\over4}} + \big({q\over4}-1\big)\gamma_{{p\over4},{q\over4}-1} - {1\over6}\big(f^{(1)}_{{1\over3},{p\over4}}f^{(1)}_{{2\over3},{q\over4}}+f^{(1)}_{{2\over3},{p\over4}}f^{(1)}_{{1\over3},{q\over4}}\big) +\frac{1}{8}\delta_{\frac{p}{4}+\frac{q}{4},1}\bigg)\nn
\end{align}
and
\be \label{f1 2 1 M3}
f^{(1)}_{{r\over3},1+{p\over4}}\big(1+{p\over4}\big)=e^w\big({p\over4} - {r\over3} \big)f^{(1)}_{{r\over3},{p\over4}}- {3\over2}f^{(1)}_{{r\over3},{1\over4}}\gamma_{{3\over4},{p\over4}} -  2f^{(1)}_{{r\over3},{1\over2}}\gamma_{{1\over 2},{p\over4}} - {3\over2}f^{(1)}_{{r\over3},{3\over4}}\gamma_{{1\over4},{p\over4}}
\ee
We note that for $N=1$ we do not get any $f^{(2)}$ terms.
The constraints (\ref{c1}) with $n=1,2$ and (\ref{c2}) become
\bea \label{gam rel M3}
{3\over2}\gamma_{{1\over4},{3\over4}} + \gamma_{{1\over2},{1\over2}}={31\over288}e^w
\eea
\bea \label{L2 M3}
\sum_{p=1}^{7}\gamma_{{p\over4},2-{p\over4}} p(2-{p\over4})= {49\over288}e^{2w}
\eea
\bea \label{L1 gam 2 1 M3}
\g_{{p\over4},{q\over4}} \!\!&=&\!\! e^{-w}{4\over p+q}\bigg(\g_{1 + {p\over4},{q\over4}}\big(1+{p\over4}\big)  + \g_{{p\over4},1+{q\over4}}\big(1+{q\over4}\big)\nn 
&&\hspace{1.8cm}+{3\over 2}\g_{{p\over4},{1\over4}}\g_{{3\over4},{q\over4}}+2\g_{{p\over4},{1\over2}}\g_{{1\over2},{q\over4}}+{3\over 2}\g_{{p\over4},{3\over4}}\g_{{1\over4},{q\over4}}\bigg)
\eea

\subsubsection*{Explicit relations}\label{subsubsection M3N1}

The set of coefficients we consider is
\bea 
\gamma_{{m\over4},{n\over4}},\quad {m\over4} + {n\over4}  \leq 2\qquad\text{and}\qquad f^{(1)}_{{r\over3},{m\over4}}, \quad {r\over3} < 1,\ {m\over4} < 2
\eea
which are the following 24 variables
\bea 
&&\g_{{1\over4},{1\over4}},\ \g_{{1\over4},{1\over2}},\ \g_{{1\over4},{3\over4}},\ \g_{{1\over4},{5\over4}},\ \g_{{1\over4},{3\over2}},\ \g_{{1\over4},{7\over4}},\ \g_{{1\over2},{1\over2}}, \ \g_{{1\over2},{3\over4}},\ \g_{{1\over2},{5\over4}},\ \g_{{1\over2},{3\over2}},\ \g_{{3\over4},{3\over4}},\ \g_{{3\over4},{5\over4}},\nn
&&f^{(1)}_{{1\over3},{1\over4}},\ f^{(1)}_{{1\over3},{1\over2}},\ f^{(1)}_{{1\over3},{3\over4}},\ f^{(1)}_{{1\over3},{5\over4}}, \ f^{(1)}_{{1\over3},{3\over2}},\ f^{(1)}_{{1\over3},{7\over4}},\ f^{(1)}_{{2\over3},{1\over4}},\ f^{(1)}_{{2\over3},{1\over2}},\ f^{(1)}_{{2\over3},{3\over4}},\ f^{(1)}_{{2\over3},{5\over4}},\ f^{(1)}_{{2\over3},{3\over2}},\ f^{(1)}_{{2\over3},{7\over4}}
\eea
At each step in section \ref{sec recursion}, we keep equations containing only the above variables. It is sufficient to proceed up to step 3) in the recursion relations. In the following analysis, we will set $w=0$, and we can reintroduce the $w$-dependence using (\ref{case 2 w}). The recursion relations are
\begin{align}\label{M3N1 recursion 1}
\hspace{-3cm}\text{step 1)}\qquad\qquad\qquad\gamma _{\frac{1}{4},\frac{1}{4}}&=-\frac{2}{3} f^{(1)}_{\frac{1}{3},\frac{1}{4}} f^{(1)}_{\frac{2}{3},\frac{1}{4}}\\[6pt]
\gamma_{\frac{1}{4},\frac{1}{2}}&=-\frac{2}{9} \left(f^{(1)}_{\frac{1}{3},\frac{1}{2}} f^{(1)}_{\frac{2}{3},\frac{1}{4}}+f^{(1)}_{\frac{1}{3},\frac{1}{4}}f^{(1)}_{\frac{2}{3},\frac{1}{2}}\right)\\[6pt]
\gamma_{\frac{1}{4},\frac{3}{4}}&=\frac{1}{6} \left(-f^{(1)}_{\frac{1}{3},\frac{3}{4}}
f^{(1)}_{\frac{2}{3},\frac{1}{4}}-f^{(1)}_{\frac{1}{3},\frac{1}{4}} f^{(1)}_{\frac{2}{3},\frac{3}{4}}\right)+\frac{1}{8}\\[6pt]
\gamma_{\frac{1}{2},\frac{1}{2}}&=\frac{1}{8}-\frac{1}{3} f^{(1)}_{\frac{1}{3},\frac{1}{2}} f^{(1)}_{\frac{2}{3},\frac{1}{2}}\\[6pt]
\gamma_{\frac{1}{2},\frac{3}{4}}&=\frac{2}{15} \left(-f^{(1)}_{\frac{1}{3},\frac{3}{4}}f^{(1)}_{\frac{2}{3},\frac{1}{2}}-f^{(1)}_{\frac{1}{3},\frac{1}{2}} f^{(1)}_{\frac{2}{3},\frac{3}{4}}\right)\\[6pt]
\gamma_{\frac{3}{4},\frac{3}{4}}&=-\frac{2}{9} f_{\frac{1}{3},\frac{3}{4}} f_{\frac{2}{3},\frac{3}{4}}
\end{align}
\begin{align}
\hspace{-1cm} \text{step 2)} \qquad\qquad \frac{5}{4} f^{(1)}_{\frac{1}{3},\frac{5}{4}}&= -\frac{1}{12}f^{(1)}_{\frac{1}{3},\frac{1}{4}}-\frac{3}{2} f^{(1)}_{\frac{1}{3},\frac{3}{4}} \gamma_{\frac{1}{4},\frac{1}{4}}-2f^{(1)}_{\frac{1}{3},\frac{1}{2}} \gamma _{\frac{1}{4},\frac{1}{2}}-\frac{3}{2} f^{(1)}_{\frac{1}{3},\frac{1}{4}}\gamma _{\frac{1}{4},\frac{3}{4}}\\[6pt]
\frac{5}{4} f^{(1)}_{\frac{2}{3},\frac{5}{4}}&=-\frac{5}{12} f^{(1)}_{\frac{2}{3},\frac{1}{4}}-
\frac{3}{2} f^{(1)}_{\frac{2}{3},\frac{3}{4}} \gamma _{\frac{1}{4},\frac{1}{4}}-2f^{(1)}_{\frac{2}{3},\frac{1}{2}} \gamma_{\frac{1}{4},\frac{1}{2}}-\frac{3}{2} f^{(1)}_{\frac{2}{3},\frac{1}{4}} \gamma _{\frac{1}{4},\frac{3}{4}}\\[6pt]
\frac{3}{2} f^{(1)}_{\frac{1}{3},\frac{3}{2}}&=\frac{1}{6}f^{(1)}_{\frac{1}{3},\frac{1}{2}}- \frac{3}{2} f^{(1)}_{\frac{1}{3},\frac{3}{4}} \gamma_{\frac{1}{4},\frac{1}{2}}-2f^{(1)}_{\frac{1}{3},\frac{1}{2}} \gamma_{\frac{1}{2},\frac{1}{2}}-\frac{3}{2} f^{(1)}_{\frac{1}{3},\frac{1}{4}}\gamma _{\frac{1}{2},\frac{3}{4}}\\[6pt] 
\frac{3}{2} f^{(1)}_{\frac{2}{3},\frac{3}{2}}&=-\frac{1}{6} f^{(1)}_{\frac{2}{3},\frac{1}{2}}-\frac{3}{2} f^{(1)}_{\frac{2}{3},\frac{3}{4}} \gamma _{\frac{1}{4},\frac{1}{2}}-2f^{(1)}_{\frac{2}{3},\frac{1}{2}} \gamma_{\frac{1}{2},\frac{1}{2}}-\frac{3}{2} f^{(1)}_{\frac{2}{3},\frac{1}{4}} \gamma _{\frac{1}{2},\frac{3}{4}}\\[6pt]
\frac{7}{4} f^{(1)}_{\frac{1}{3},\frac{7}{4}}&=\frac{5}{12}f^{(1)}_{\frac{1}{3},\frac{3}{4}}- \frac{3}{2} f^{(1)}_{\frac{1}{3},\frac{3}{4}} \gamma_{\frac{1}{4},\frac{3}{4}}-2f^{(1)}_{\frac{1}{3},\frac{1}{2}} \gamma _{\frac{1}{2},\frac{3}{4}}-\frac{3}{2} f^{(1)}_{\frac{1}{3},\frac{1}{4}}
\gamma_{\frac{3}{4},\frac{3}{4}}\\[6pt]
\frac{7}{4} f^{(1)}_{\frac{2}{3},\frac{7}{4}}&=\frac{1}{12} f^{(1)}_{\frac{2}{3},\frac{3}{4}}-
\frac{3}{2} f^{(1)}_{\frac{2}{3},\frac{3}{4}} \gamma _{\frac{1}{4},\frac{3}{4}}-2f^{(1)}_{\frac{2}{3},\frac{1}{2}} \gamma_{\frac{1}{2},\frac{3}{4}}-\frac{3}{2} f^{(1)}_{\frac{2}{3},\frac{1}{4}} \gamma _{\frac{3}{4},\frac{3}{4}}
\end{align}
\begin{align}
\hspace{-2cm}\text{step 3)}\qquad\qquad\quad\gamma_{\frac{1}{4},\frac{5}{4}}&=\frac{2}{3} \left(-\frac{1}{6} \left(f^{(1)}_{\frac{1}{3},\frac{5}{4}}
f^{(1)}_{\frac{2}{3},\frac{1}{4}}+f^{(1)}_{\frac{1}{3},\frac{1}{4}} f^{(1)}_{\frac{2}{3},\frac{5}{4}}\right)+\frac{1}{4} \gamma_{\frac{1}{4},\frac{1}{4}}\right)\\[6pt]
\gamma_{\frac{1}{4},\frac{3}{2}}&=\frac{4}{7} \left(-\frac{1}{6}\left(f^{(1)}_{\frac{1}{3},\frac{3}{2}} f^{(1)}_{\frac{2}{3},\frac{1}{4}}+f^{(1)}_{\frac{1}{3},\frac{1}{4}}f^{(1)}_{\frac{2}{3},\frac{3}{2}}\right)+\frac{1}{2} \gamma _{\frac{1}{4},\frac{1}{2}}\right)\\[6pt]
\gamma_{\frac{1}{4},\frac{7}{4}}&=\frac{1}{2} \left(-\frac{1}{6} \left(f^{(1)}_{\frac{1}{3},\frac{7}{4}}
f^{(1)}_{\frac{2}{3},\frac{1}{4}}+f^{(1)}_{\frac{1}{3},\frac{1}{4}} f^{(1)}_{\frac{2}{3},\frac{7}{4}}\right)+\frac{3}{4} \gamma_{\frac{1}{4},\frac{3}{4}}\right)\\[6pt]
\gamma _{\frac{1}{2},\frac{5}{4}}&=\frac{4}{7} \left(-\frac{1}{6}\left(f^{(1)}_{\frac{1}{3},\frac{5}{4}} f^{(1)}_{\frac{2}{3},\frac{1}{2}}+f^{(1)}_{\frac{1}{3},\frac{1}{2}}f^{(1)}_{\frac{2}{3},\frac{5}{4}}\right)+\frac{1}{4} \gamma _{\frac{1}{4},\frac{1}{2}}\right)\\[6pt]
\gamma_{\frac{1}{2},\frac{3}{2}}&=\frac{1}{2} \left(-\frac{1}{6} \left(f^{(1)}_{\frac{1}{3},\frac{3}{2}}
f^{(1)}_{\frac{2}{3},\frac{1}{2}}+f^{(1)}_{\frac{1}{3},\frac{1}{2}} f^{(1)}_{\frac{2}{3},\frac{3}{2}}\right)+\frac{1}{2} \gamma_{\frac{1}{2},\frac{1}{2}}\right)\\[6pt]
\label{M3N1 recursion 18}
\gamma _{\frac{3}{4},\frac{5}{4}}&=\frac{1}{2} \left(-\frac{1}{6}
\left(f^{(1)}_{\frac{1}{3},\frac{5}{4}} f^{(1)}_{\frac{2}{3},\frac{3}{4}}+f^{(1)}_{\frac{1}{3},\frac{3}{4}}f^{(1)}_{\frac{2}{3},\frac{5}{4}}\right)+\frac{1}{4} \gamma_{\frac{1}{4},\frac{3}{4}}\right)
\end{align}
The constraints are
\begin{align}\label{M3N1 constraint 1}
\frac{31}{288}&=\gamma_{\frac{1}{2},\frac{1}{2}}+\frac{3}{2} \gamma _{\frac{1}{4},\frac{3}{4}}\\[6pt]
\label{M3N1 constraint 6}
\frac{49}{288}&=\frac{15}{2} \gamma_{\frac{3}{4},\frac{5}{4}}+6 \gamma _{\frac{1}{2},\frac{3}{2}}+\frac{7}{2} \gamma _{\frac{1}{4},\frac{7}{4}}\\[6pt]
\gamma _{\frac{1}{4},\frac{1}{4}}&=2 \left(2\gamma _{\frac{1}{4},\frac{1}{2}}^2+3 \gamma _{\frac{1}{4},\frac{1}{4}} \gamma_{\frac{1}{4},\frac{3}{4}}+\frac{5}{2} \gamma _{\frac{1}{4},\frac{5}{4}}\right)\\[6pt]
\gamma_{\frac{1}{4},\frac{1}{2}}&=\frac{4}{3} \left(2\gamma _{\frac{1}{4},\frac{1}{2}} \gamma_{\frac{1}{2},\frac{1}{2}}+\frac{3}{2} \gamma _{\frac{1}{4},\frac{1}{2}} \gamma _{\frac{1}{4},\frac{3}{4}}+\frac{3}{2} \gamma_{\frac{1}{4},\frac{1}{4}} \gamma _{\frac{1}{2},\frac{3}{4}}+\frac{5}{4} \gamma _{\frac{1}{2},\frac{5}{4}}+\frac{3}{2}\gamma _{\frac{1}{4},\frac{3}{2}}\right)\\[6pt]
\gamma _{\frac{1}{2},\frac{1}{2}}&=2\gamma _{\frac{1}{2},\frac{1}{2}}^2+3\gamma _{\frac{1}{4},\frac{1}{2}} \gamma _{\frac{1}{2},\frac{3}{4}}+3 \gamma _{\frac{1}{2},\frac{3}{2}}\\[6pt]
\gamma_{\frac{1}{4},\frac{3}{4}}&=\frac{3}{2} \gamma _{\frac{1}{4},\frac{3}{4}}^2+2\gamma _{\frac{1}{4},\frac{1}{2}} \gamma_{\frac{1}{2},\frac{3}{4}}+\frac{3}{2} \gamma _{\frac{1}{4},\frac{1}{4}} \gamma _{\frac{3}{4},\frac{3}{4}}+\frac{5}{4}\gamma _{\frac{3}{4},\frac{5}{4}}+\frac{7}{4} \gamma _{\frac{1}{4},\frac{7}{4}}
\end{align}
By using the recursion relations, all 24 variables can be determined using only 6 variables, as indicated in (\ref{inputs})
\bea 
f^{(1)}_{{1\over3},{1\over4}},\ f^{(1)}_{{1\over3},{1\over2}},\  f^{(1)}_{{1\over3},{3\over4}},\ f^{(1)}_{{2\over3},{1\over4}},\ f^{(1)}_{{2\over3},{1\over2}},\ f^{(1)}_{{2\over3},{3\over4}}
\eea
Inserting the recursion relations (\ref{M3N1 recursion 1} -- \ref{M3N1 recursion 18}) into the 6 constraints, we find that 4 of them turn out to be trivial, while 2 constraints (\ref{M3N1 constraint 1}) and (\ref{M3N1 constraint 6}) remain nontrivial. These nontrivial constraints are
\begin{align}
\frac{31}{288}&=-\frac{1}{4} f^{(1)}_{\frac{1}{3},\frac{3}{4}} f^{(1)}_{\frac{2}{3},\frac{1}{4}}-\frac{1}{3}
   f^{(1)}_{\frac{1}{3},\frac{1}{2}} f^{(1)}_{\frac{2}{3},\frac{1}{2}}-\frac{1}{4} f^{(1)}_{\frac{1}{3},\frac{1}{4}}
   f^{(1)}_{\frac{2}{3},\frac{3}{4}}+\frac{5}{16} 
   \\
   \frac{49}{288}&=-\frac{1}{6}
   (f^{(1)}_{\frac{1}{3},\frac{3}{4}})^2 (f^{(1)}_{\frac{2}{3},\frac{1}{4}})^2-\frac{4}{9}
   (f^{(1)}_{\frac{1}{3},\frac{1}{2}})^2 (f^{(1)}_{\frac{2}{3},\frac{1}{2}})^2-\frac{1}{6}
   (f^{(1)}_{\frac{1}{3},\frac{1}{4}})^2 (f^{(1)}_{\frac{2}{3},\frac{3}{4}})^2+\frac{1}{6}
   f^{(1)}_{\frac{1}{3},\frac{1}{2}} f^{(1)}_{\frac{2}{3},\frac{1}{2}}\nn 
   &+\frac{3}{2} \left(\frac{1}{8}-\frac{1}{3}
   f^{(1)}_{\frac{1}{3},\frac{1}{2}} f^{(1)}_{\frac{2}{3},\frac{1}{2}}\right)+f^{(1)}_{\frac{2}{3},\frac{3}{4}}\left(-\frac{4}{9}
   f^{(1)}_{\frac{2}{3},\frac{1}{4}} (f^{(1)}_{\frac{1}{3},\frac{1}{2}})^2-\frac{4}{9} f^{(1)}_{\frac{1}{3},\frac{1}{4}}
   f^{(1)}_{\frac{2}{3},\frac{1}{2}} f^{(1)}_{\frac{1}{3},\frac{1}{2}}-\frac{2}{9}
   f^{(1)}_{\frac{1}{3},\frac{1}{4}}\right) \nn 
   &+f^{(1)}_{\frac{1}{3},\frac{3}{4}}
   \left(-\frac{4}{9} f^{(1)}_{\frac{1}{3},\frac{1}{4}} (f^{(1)}_{\frac{2}{3},\frac{1}{2}})^2-\frac{4}{9}
   f^{(1)}_{\frac{1}{3},\frac{1}{2}} f^{(1)}_{\frac{2}{3},\frac{1}{4}} f^{(1)}_{\frac{2}{3},\frac{1}{2}}-\frac{1}{9}
   f^{(1)}_{\frac{2}{3},\frac{1}{4}}\right)
   -\frac{13}{9}f^{(1)}_{\frac{1}{3},\frac{3}{4}}f^{(1)}_{\frac{2}{3},\frac{3}{4}}
    f^{(1)}_{\frac{1}{3},\frac{1}{4}} f^{(1)}_{\frac{2}{3},\frac{1}{4}}
   +\frac{9}{32}
\end{align}
We can solve these constraints in terms of only four variables
\bea \label{M3N1 variables}
f^{(1)}_{{1\over3},{1\over4}},\ f^{(1)}_{{1\over3},{1\over2}},\ f^{(1)}_{{2\over3},{1\over4}},\ f^{(1)}_{{2\over3},{1\over2}}
\eea
as 
\begin{align}\label{f 1over3 3over4}
    f^{(1)}_{\frac{1}{3},\frac{3}{4}}&={1\over720 (f_{\frac{1}{3},\frac{1}{4}}^{(1)})
   (f_{\frac{2}{3},\frac{1}{4}}^{(1)}){}^2} \bigg[2 \bigg(64512 \big(f_{\frac{1}{3},\frac{1}{2}}^{(1)}\big){}^2
   \big(f_{\frac{1}{3},\frac{1}{4}}^{(1)}\big){}^2 \big(f_{\frac{2}{3},\frac{1}{2}}^{(1)}\big){}^2
   \big(f_{\frac{2}{3},\frac{1}{4}}^{(1)}\big){}^2\nn 
   &-34560 \big(f_{\frac{1}{3},\frac{1}{2}}^{(1)}\big){}^3
   \big(f_{\frac{1}{3},\frac{1}{4}}^{(1)}\big) \big(f_{\frac{2}{3},\frac{1}{2}}^{(1)}\big)
   \big(f_{\frac{2}{3},\frac{1}{4}}^{(1)}\big){}^3+23832 \big(f_{\frac{1}{3},\frac{1}{2}}^{(1)}\big){}^2
   \big(f_{\frac{1}{3},\frac{1}{4}}^{(1)}\big) \big(f_{\frac{2}{3},\frac{1}{4}}^{(1)}\big){}^3+5184
   \big(f_{\frac{1}{3},\frac{1}{2}}^{(1)}\big){}^4 \big(f_{\frac{2}{3},\frac{1}{4}}^{(1)}\big){}^4\nn 
   &-34560\big(f_{\frac{1}{3},\frac{1}{2}}^{(1)}\big) \big(f_{\frac{1}{3},\frac{1}{4}}^{(1)}\big){}^3
   \big(f_{\frac{2}{3},\frac{1}{2}}^{(1)}\big){}^3 \big(f_{\frac{2}{3},\frac{1}{4}}^{(1)}\big)-57840
   \big(f_{\frac{1}{3},\frac{1}{2}}^{(1)}\big) \big(f_{\frac{1}{3},\frac{1}{4}}^{(1)}\big){}^2 \big(f_{\frac{2}{3},\frac{1}{2}}^{(1)}\big)
   \big(f_{\frac{2}{3},\frac{1}{4}}^{(1)}\big){}^2\nn 
   &+18648 \big(f_{\frac{1}{3},\frac{1}{4}}^{(1)}\big){}^3
   \big(f_{\frac{2}{3},\frac{1}{2}}^{(1)}\big){}^2 \big(f_{\frac{2}{3},\frac{1}{4}}^{(1)}\big)+5184
   \big(f_{\frac{1}{3},\frac{1}{4}}^{(1)}\big){}^4 \big(f_{\frac{2}{3},\frac{1}{2}}^{(1)}\big){}^4+16234
   \big(f_{\frac{1}{3},\frac{1}{4}}^{(1)}\big){}^2 \big(f_{\frac{2}{3},\frac{1}{4}}^{(1)}\big){}^2\bigg)^{1\over2}\nn 
   &-144\big(f_{\frac{1}{3},\frac{1}{2}}^{(1)}\big){}^2 \big(f_{\frac{2}{3},\frac{1}{4}}^{(1)}\big){}^2-480
   \big(f_{\frac{1}{3},\frac{1}{2}}^{(1)}\big) \big(f_{\frac{1}{3},\frac{1}{4}}^{(1)}\big) \big(f_{\frac{2}{3},\frac{1}{2}}^{(1)}\big)
   \big(f_{\frac{2}{3},\frac{1}{4}}^{(1)}\big)+144 \big(f_{\frac{1}{3},\frac{1}{4}}^{(1)}\big){}^2
   \big(f_{\frac{2}{3},\frac{1}{2}}^{(1)}\big){}^2\nn 
   &+259 \big(f_{\frac{1}{3},\frac{1}{4}}^{(1)}\big)
   \big(f_{\frac{2}{3},\frac{1}{4}}^{(1)}\big)\bigg]
   \\
  \label{f 2over3 3over4}
   f^{(1)}_{\frac{2}{3},\frac{3}{4}}&={1\over720 \left(f_{\frac{1}{3},\frac{1}{4}}^{(1)}\right){}^2
   \left(f_{\frac{2}{3},\frac{1}{4}}^{(1)}\right)} \bigg[-2 \bigg(64512 \big(f_{\frac{1}{3},\frac{1}{2}}^{(1)}\big){}^2
   \big(f_{\frac{1}{3},\frac{1}{4}}^{(1)}\big){}^2 \big(f_{\frac{2}{3},\frac{1}{2}}^{(1)}\big){}^2
   \big(f_{\frac{2}{3},\frac{1}{4}}^{(1)}\big){}^2\nn 
   &-34560 \big(f_{\frac{1}{3},\frac{1}{2}}^{(1)}\big){}^3
   \big(f_{\frac{1}{3},\frac{1}{4}}^{(1)}\big) \big(f_{\frac{2}{3},\frac{1}{2}}^{(1)}\big)
   \big(f_{\frac{2}{3},\frac{1}{4}}^{(1)}\big){}^3+23832 \big(f_{\frac{1}{3},\frac{1}{2}}^{(1)}\big){}^2
   \big(f_{\frac{1}{3},\frac{1}{4}}^{(1)}\big) \big(f_{\frac{2}{3},\frac{1}{4}}^{(1)}\big){}^3+5184
   \big(f_{\frac{1}{3},\frac{1}{2}}^{(1)}\big){}^4 \big(f_{\frac{2}{3},\frac{1}{4}}^{(1)}\big){}^4\nn 
   &-34560\big(f_{\frac{1}{3},\frac{1}{2}}^{(1)}\big) \big(f_{\frac{1}{3},\frac{1}{4}}^{(1)}\big){}^3
   \big(f_{\frac{2}{3},\frac{1}{2}}^{(1)}\big){}^3 \big(f_{\frac{2}{3},\frac{1}{4}}^{(1)}\big)-57840
   \big(f_{\frac{1}{3},\frac{1}{2}}^{(1)}\big) \big(f_{\frac{1}{3},\frac{1}{4}}^{(1)}\big){}^2 \big(f_{\frac{2}{3},\frac{1}{2}}^{(1)}\big)
   \big(f_{\frac{2}{3},\frac{1}{4}}^{(1)}\big){}^2\nn 
   &+18648 \big(f_{\frac{1}{3},\frac{1}{4}}^{(1)}\big){}^3
   \big(f_{\frac{2}{3},\frac{1}{2}}^{(1)}\big){}^2 \big(f_{\frac{2}{3},\frac{1}{4}}^{(1)}\big)+5184
   \big(f_{\frac{1}{3},\frac{1}{4}}^{(1)}\big){}^4 \big(f_{\frac{2}{3},\frac{1}{2}}^{(1)}\big){}^4+16234
   \big(f_{\frac{1}{3},\frac{1}{4}}^{(1)}\big){}^2 \big(f_{\frac{2}{3},\frac{1}{4}}^{(1)}\big){}^2\bigg)^{1\over2}\nn
   &+144
   \big(f_{\frac{1}{3},\frac{1}{2}}^{(1)}\big){}^2 \big(f_{\frac{2}{3},\frac{1}{4}}^{(1)}\big){}^2-480
   \big(f_{\frac{1}{3},\frac{1}{2}}^{(1)}\big) \big(f_{\frac{1}{3},\frac{1}{4}}^{(1)}\big) \big(f_{\frac{2}{3},\frac{1}{2}}^{(1)}\big)
   \big(f_{\frac{2}{3},\frac{1}{4}}^{(1)}\big)-144 \big(f_{\frac{1}{3},\frac{1}{4}}^{(1)}\big){}^2
   \big(f_{\frac{2}{3},\frac{1}{2}}^{(1)}\big){}^2\nn 
   &+331 \big(f_{\frac{1}{3},\frac{1}{4}}^{(1)}\big)
   \big(f_{\frac{2}{3},\frac{1}{4}}^{(1)}\big)\bigg]
\end{align}
Using these and the recursion relations, all 10 variables can be determined completely in terms of the 4 variables (\ref{M3N1 variables}). The values for these inputs are given by (\ref{f1}) which we write below 
\begin{align}
\label{input M3N1}
f^{(1)}_{{1\over3},{1\over4}}&={3^{1\over4}\over\sqrt2(-1+i)^{1\over3}} & f^{(1)}_{{1\over3},{1\over2}}&={(-1+i)^{2\over3}\over2\sqrt3}\nn 
f^{(1)}_{{2\over3},{1\over4}}&={({1\over4}-{i\over4})3^{1\over4}\over2^{5\over6}}&f^{(1)}_{{2\over3},{1\over2}}&=-{5i\over4\ 2^{1\over3}\sqrt3} 
\end{align}
Inserting the values (\ref{input M3N1}) into the system of equations (\ref{f 1over3 3over4}), (\ref{f 2over3 3over4}), and (\ref{M3N1 recursion 1}~--~\ref{M3N1 recursion 18}),
we obtain the following values for the remaining variables,

\allowdisplaybreaks
\begin{align}
\gamma_{\frac{1}{4},\frac{1}{4}}&= \frac{i}{4 \sqrt{3}},&\gamma_{\frac{1}{4},\frac{1}{2}}&=\frac{(-\frac{1}{3})^{1\over4}}{9} ,&\gamma_{\frac{1}{4},\frac{3}{4}}&=\frac{5}{144},&\gamma_{\frac{1}{4},\frac{5}{4}}&=\frac{77 i}{2592 \sqrt{3}},
\nn 
\gamma_{\frac{1}{4},\frac{3}{2}}&= \frac{2(-\frac{1}{3})^{1\over4}}{81},&\gamma_{\frac{1}{4},\frac{7}{4}}&=\frac{221}{20736},&\gamma_{\frac{1}{2},\frac{1}{2}}&=\frac{1}{18},&\gamma_{\frac{1}{2},\frac{5}{4}}&=\frac{11(-\frac{1}{3})^{1\over4}}{648},
   \nn
   \gamma_{\frac{3}{4},\frac{5}{4}}&=\frac{385}{62208},&\gamma_{\frac{1}{2},\frac{3}{4}}&= -\frac{(-\frac{1}{3})^{3\over4}}{18} ,&\gamma_{\frac{3}{4},\frac{3}{4}}&= -\frac{25 i}{1296 \sqrt{3}},&\gamma_{\frac{1}{2},\frac{3}{2}}&=\frac{7}{486},
   \nn 
   f_{\frac{1}{3},\frac{3}{4}}&= \frac{(-\frac{1}{3})^{1\over4}}{6\ 2^{2\over3}},&f_{\frac{2}{3},\frac{3}{4}}&=\frac{25 (-\frac{1}{3})^{1\over4}}{24\ 2^{1\over3}},&f_{\frac{1}{3},\frac{5}{4}}&=\frac{7(-{1\over3})^{3\over4}}{72\ 2^{2\over3} },&f_{\frac{2}{3},\frac{5}{4}}&=\frac{55(-{1\over3})^{3\over4} }{288\  2^{1\over3}}, 
   \nn
f_{\frac{1}{3},\frac{3}{2}}&= \frac{i 2^{1\over3}}{27 \sqrt{3}},&f_{\frac{2}{3},\frac{3}{2}}&=\frac{7 i}{54\ 2^{1\over3} \sqrt{3}},&f_{\frac{1}{3},\frac{7}{4}}&=
   \frac{13(-\frac{1}{3})^{1\over4}}{432\ 2^{2\over3}},&f_{\frac{2}{3},\frac{7}{4}}&= \frac{85 (-\frac{1}{3})^{1\over4}}{1728\ 2^{1\over3}}
\end{align}
which agree with the known results from (\ref{MN gamma}) and (\ref{f1}) again demonstrating that the bootstrap method generates the correct equations. 
Note that the number of inputs is 4, in agreement with equation (\ref{N inputs}) for $(M,N)=(3,1)$ and $N^{\text{constr}}=2$. 

\subsection{$M=N=2$}

In this case, the $w$-dependence is given by
\bea\label{case 3 w}
\gamma_{{p\over4},{p\over4}}&\propto&e^{{p+q\over 4} w}\nn
f^{(1)}_{{r\over2},{p\over4}}&\propto& e^{({p\over4}-{r\over2})w}\nn
f^{(2)}_{{r\over2},{p\over4}}&\propto& e^{({p\over4}-{r\over2})w}
\eea
The recursion relations (\ref{r1}), (\ref{r2}) and (\ref{r3}) are given by
\begin{align} \label{gam greater one M2N2}
\gamma_{{p\over4},{q\over4}}&=e^w{4\over p+q}\bigg(\big({p\over4}-1\big)\gamma_{{p\over4}-1,{q\over4}}+ \big({q\over4}-1\big)\gamma_{{p\over4},{q\over4}-1}- {1\over4}f^{(1)}_{{1\over2},{p\over4}}f^{(1)}_{{1\over2},{q\over4}} - {1\over4}f^{(2)}_{{1\over2},{p\over4}}f^{(2)}_{{1\over2},{q\over4}}+\frac{1}{8}\delta_{\frac{p}{4}+\frac{q}{4},1}\bigg)\nn
\end{align} 
and 
\be\label{f1 2 M=N=2}
   f^{(i)}_{\frac{r}{2},\frac{p}{4}+1}\big(\frac{p}{4}+1\big)=e^w\big(\frac{p}{4}-\frac{r}{2}\big) f^{(i)}_{\frac{r}{2},\frac{p}{4}} -\frac{3}{2} f^{(i)}_{\frac{r}{2},\frac{3}{4}}
   \gamma _{\frac{1}{4},\frac{p}{4}}-2f^{(i)}_{\frac{r}{2},\frac{1}{2}} \gamma _{\frac{1}{2},\frac{p}{4}}-\frac{3}{2}
   f^{(i)}_{\frac{r}{2},\frac{1}{4}} \gamma _{\frac{3}{4},\frac{p}{4}}
\ee
where $i=1,2$.

The constraints (\ref{c1}) with $n=1,2$ and (\ref{c2}) become 
\be\label{gamma dimension relation 2 M=N=2}
 \gamma_{\frac{1}{2},\frac{1}{2}}+\frac{3}{2} \gamma _{\frac{3}{4},\frac{1}{4}}=\frac{3}{32}e^{w}
\ee
\be \label{L2 M=N=2}
\sum_{p=1}^{7}\gamma_{{p\over4},2-{p\over4}}p(2-{p\over4})= {5\over32}e^{2w}
\ee
\bea \label{g 2 M=N=2}
\gamma _{\frac{p}{4},\frac{q}{4}}&=&e^{-w}\frac{4}{p+q} \bigg(\left(\frac{p}{4}+1\right) \gamma _{\frac{p}{4}+1,\frac{q}{4}}+\left(\frac{q}{4}+1\right) \gamma
   _{\frac{p}{4},\frac{q}{4}+1}\nn 
   &&+\frac{3}{2} \gamma _{\frac{p}{4},\frac{1}{4}} \gamma _{\frac{3}{4},\frac{q}{4}}+2\gamma
   _{\frac{p}{4},\frac{1}{2}} \gamma _{\frac{1}{2},\frac{q}{4}}+\frac{3}{2} \gamma _{\frac{p}{4},\frac{3}{4}}
   \gamma _{\frac{1}{4},\frac{q}{4}}\bigg)
\eea

\subsubsection*{Explicit relations}\label{subsubsection M2N2}

The set of coefficients we consider is
\bea 
\gamma_{{m\over4},{n\over4}},\quad {m\over4}+{n\over4}\leq 2\qquad\text{and}\qquad 
f^{(i)}_{{r\over2},{m\over4}},\quad {r\over2} < 1,\ {m\over4} < 2,\quad i=1,2
\eea
which are the following 24 variables
\bea
&&\gamma_{\frac{1}{4},\frac{1}{4}},\ \gamma_{\frac{1}{4},\frac{1}{2}},\ \gamma_{\frac{1}{4},\frac{3}{4}},\ \gamma_{\frac{1}{4},\frac{5}{4}},\ \gamma_{\frac{1}{4},\frac{3}{2}},\ \gamma_{\frac{1}{4},\frac{7}{4}},\ \gamma_{\frac{1}{2},\frac{1}{2}},\ \gamma_{\frac{1}{2},\frac{3}{4}},\ \gamma_{\frac{1}{2},\frac{5}{4}},\ \gamma_{\frac{1}{2},\frac{3}{2}},\ \gamma_{\frac{3}{4},\frac{3}{4}}, \gamma_{\frac{3}{4},\frac{5}{4}}
\nn
&&f^{(1)}_{\frac{1}{2},\frac{1}{4}},\ f^{(1)}_{\frac{1}{2},\frac{1}{2}},\ f^{(1)}_{\frac{1}{2},\frac{3}{4}},\ f^{(1)}_{\frac{1}{2},\frac{5}{4}},\ f^{(1)}_{\frac{1}{2},\frac{3}{2}},\ f^{(1)}_{\frac{1}{2},\frac{7}{4}},\ f^{(2)}_{\frac{1}{2},\frac{1}{4}},\ f^{(2)}_{\frac{1}{2},\frac{1}{2}},\ f^{(2)}_{\frac{1}{2},\frac{3}{4}},\ f^{(2)}_{\frac{1}{2},\frac{5}{4}},\ f^{(2)}_{\frac{1}{2},\frac{3}{2}},\ f^{(2)}_{\frac{1}{2},\frac{7}{4}}
\eea 
At each step in section \ref{sec recursion}, we keep equations containing only the above variables. It is sufficient to proceed up to step 3) in the recursion relations. In the following analysis, we will again set $w=0$, and we can reintroduce the $w$-dependence using (\ref{case 3 w}). The recursion relations are 
\begin{align}\label{M2N2 recursion relation 1}
\hspace{-3cm}\text{step 1)}\qquad\qquad\qquad\qquad\gamma_{\frac{1}{4},\frac{1}{4}}&=-2 \left(\frac{1}{4} \big(f^{(1)}_{\frac{1}{2},\frac{1}{4}}\big)^2+\frac{1}{4}
   \big(f^{(2)}_{\frac{1}{2},\frac{1}{4}}\big)^2\right)\\[6pt]
\gamma_{\frac{1}{4},\frac{1}{2}}&=-\frac{4}{3} \left(\frac{1}{4}
   f^{(1)}_{\frac{1}{2},\frac{1}{4}} f^{(1)}_{\frac{1}{2},\frac{1}{2}}+\frac{1}{4} f^{(2)}_{\frac{1}{2},\frac{1}{4}}f^{(2)}_{\frac{1}{2},\frac{1}{2}}\right)\\[6pt]   
\gamma_{\frac{1}{4},\frac{3}{4}}&=-\frac{1}{4} f^{(1)}_{\frac{1}{2},\frac{1}{4}}
   f^{(1)}_{\frac{1}{2},\frac{3}{4}}-\frac{1}{4} f^{(2)}_{\frac{1}{2},\frac{1}{4}} f^{(2)}_{\frac{1}{2},\frac{3}{4}}+\frac{1}{8}\\[6pt]
\gamma_{\frac{1}{2},\frac{1}{2}}&=-\frac{1}{4} \big(f^{(1)}_{\frac{1}{2},\frac{1}{2}}\big)^2-\frac{1}{4}
   \big(f^{(2)}_{\frac{1}{2},\frac{1}{2}}\big)^2+\frac{1}{8}\\[6pt]
\gamma_{\frac{1}{2},\frac{3}{4}}&=\frac{4}{5} \left(-\frac{1}{4} f^{(1)}_{\frac{1}{2},\frac{1}{2}} 
   f^{(1)}_{\frac{1}{2},\frac{3}{4}}-\frac{1}{4}
   f^{(2)}_{\frac{1}{2},\frac{1}{2}} f^{(2)}_{\frac{1}{2},\frac{3}{4}}\right)\\[6pt]
\gamma_{\frac{3}{4},\frac{3}{4}}&=\frac{2}{3} \left(-\frac{1}{4} \big(f^{(1)}_{\frac{1}{2},\frac{3} 
   {4}}\big)^2-\frac{1}{4}\big(f^{(2)}_{\frac{1}{2},\frac{3}{4}}\big)^2\right)
   \end{align}
\begin{align}
\hspace{-2cm}\text{step 2)}\qquad\qquad\frac{5}{4} f^{(1)}_{\frac{1}{2},\frac{5}{4}}&=-\frac{1}{4} f^{(1)}_{\frac{1}{2},\frac{1}{4}}-
   \frac{3}{2} f^{(1)}_{\frac{1}{2},\frac{3}{4}} \gamma _{\frac{1}{4},\frac{1}{4}}-2f^{(1)}_{\frac{1}{2},\frac{1}{2}} \gamma
   _{\frac{1}{4},\frac{1}{2}}-\frac{3}{2} f^{(1)}_{\frac{1}{2},\frac{1}{4}} \gamma
   _{\frac{1}{4},\frac{3}{4}}\\[6pt]
\frac{3}{2} f^{(1)}_{\frac{1}{2},\frac{3}{2}}&=-\frac{3}{2}
   f^{(1)}_{\frac{1}{2},\frac{3}{4}} \gamma _{\frac{1}{4},\frac{1}{2}}-2f^{(1)}_{\frac{1}{2},\frac{1}{2}} \gamma_{\frac{1}{2},\frac{1}{2}}-\frac{3}{2} f^{(1)}_{\frac{1}{2},\frac{1}{4}} \gamma_{\frac{1}{2},\frac{3}{4}}\\[6pt]  
\frac{7}{4} f^{(1)}_{\frac{1}{2},\frac{7}{4}}&=\frac{1}{4} f^{(1)}_{\frac{1}{2},\frac{3}{4}}-\frac{3}{2} f^{(1)}_{\frac{1}{2},\frac{3}{4}} \gamma _{\frac{1}{4},\frac{3}{4}}-2f^{(1)}_{\frac{1}{2},\frac{1}{2}} \gamma_{\frac{1}{2},\frac{3}{4}}-\frac{3}{2} f^{(1)}_{\frac{1}{2},\frac{1}{4}} \gamma_{\frac{3}{4},\frac{3}{4}}\\[6pt]
\frac{5}{4} f^{(2)}_{\frac{1}{2},\frac{5}{4}}&=-\frac{1}{4} f^{(2)}_{\frac{1}{2},\frac{1}{4}}-\frac{3}{2} f^{(2)}_{\frac{1}{2},\frac{3}{4}} \gamma _{\frac{1}{4},\frac{1}{4}}-2f^{(2)}_{\frac{1}{2},\frac{1}{2}} \gamma
   _{\frac{1}{4},\frac{1}{2}}-\frac{3}{2} f^{(2)}_{\frac{1}{2},\frac{1}{4}} \gamma
   _{\frac{1}{4},\frac{3}{4}}\\[6pt]
\frac{3}{2} f^{(2)}_{\frac{1}{2},\frac{3}{2}}&=-\frac{3}{2}
   f^{(2)}_{\frac{1}{2},\frac{3}{4}} \gamma _{\frac{1}{4},\frac{1}{2}}-2f^{(2)}_{\frac{1}{2},\frac{1}{2}} \gamma_{\frac{1}{2},\frac{1}{2}}-\frac{3}{2} f^{(2)}_{\frac{1}{2},\frac{1}{4}} \gamma_{\frac{1}{2},\frac{3}{4}}\\[6pt]
\frac{7}{4} f^{(2)}_{\frac{1}{2},\frac{7}{4}}&=\frac{1}{4} f^{(2)}_{\frac{1}{2},\frac{3}{4}}-\frac{3}{2} f^{(2)}_{\frac{1}{2},\frac{3}{4}} \gamma _{\frac{1}{4},\frac{3}{4}}-2f^{(2)}_{\frac{1}{2},\frac{1}{2}} \gamma_{\frac{1}{2},\frac{3}{4}}-\frac{3}{2} f^{(2)}_{\frac{1}{2},\frac{1}{4}} \gamma_{\frac{3}{4},\frac{3}{4}}
\end{align}
\begin{align}
\hspace{-2.5cm}\text{step 3)}\qquad\qquad\qquad \gamma_{\frac{1}{4},\frac{5}{4}}&=\frac{2}{3} \left(-\frac{1}{4}
   f^{(1)}_{\frac{1}{2},\frac{1}{4}} f^{(1)}_{\frac{1}{2},\frac{5}{4}}-\frac{1}{4} f^{(2)}_{\frac{1}{2},\frac{1}{4}}f^{(2)}_{\frac{1}{2},\frac{5}{4}}+\frac{1}{4} \gamma _{\frac{1}{4},\frac{1}{4}}\right)\\[6pt]  
\gamma_{\frac{1}{4},\frac{3}{2}}&=\frac{4}{7} \left(-\frac{1}{4} f^{(1)}_{\frac{1}{2},\frac{1}{4}}
   f^{(1)}_{\frac{1}{2},\frac{3}{2}}-\frac{1}{4} f^{(2)}_{\frac{1}{2},\frac{1}{4}} f^{(2)}_{\frac{1}{2},\frac{3}{2}}+\frac{1}{2} \gamma_{\frac{1}{4},\frac{1}{2}}\right)\\[6pt]
\gamma_{\frac{1}{4},\frac{7}{4}}&=\frac{1}{2} \left(-\frac{1}{4}
   f^{(1)}_{\frac{1}{2},\frac{1}{4}} f^{(1)}_{\frac{1}{2},\frac{7}{4}}-\frac{1}{4} f^{(2)}_{\frac{1}{2},\frac{1}{4}}f^{(2)}_{\frac{1}{2},\frac{7}{4}}+\frac{3}{4} \gamma _{\frac{1}{4},\frac{3}{4}}\right)\\[6pt] 
\gamma_{\frac{1}{2},\frac{5}{4}}&=\frac{4}{7} \left(-\frac{1}{4} f^{(1)}_{\frac{1}{2},\frac{1}{2}}
   f^{(1)}_{\frac{1}{2},\frac{5}{4}}-\frac{1}{4} f^{(2)}_{\frac{1}{2},\frac{1}{2}} f^{(2)}_{\frac{1}{2},\frac{5}{4}}+\frac{1}{4} \gamma
   _{\frac{1}{4},\frac{1}{2}}\right)\\[6pt]
\gamma_{\frac{1}{2},\frac{3}{2}}&=\frac{1}{2} \left(-\frac{1}{4}
   f^{(1)}_{\frac{1}{2},\frac{1}{2}} f^{(1)}_{\frac{1}{2},\frac{3}{2}}-\frac{1}{4} f^{(2)}_{\frac{1}{2},\frac{1}{2}}
   f^{(2)}_{\frac{1}{2},\frac{3}{2}}+\frac{1}{2} \gamma _{\frac{1}{2},\frac{1}{2}}\right)\\[6pt]
   \label{M2N2 recursion relation 18}
\gamma_{\frac{3}{4},\frac{5}{4}}&=\frac{1}{2} \left(-\frac{1}{4} f^{(1)}_{\frac{1}{2},\frac{3}{4}}
   f^{(1)}_{\frac{1}{2},\frac{5}{4}}-\frac{1}{4} f^{(2)}_{\frac{1}{2},\frac{3}{4}} f^{(2)}_{\frac{1}{2},\frac{5}{4}}+\frac{1}{4} \gamma_{\frac{1}{4},\frac{3}{4}}\right)
\end{align}
The constraints are
\begin{align}
\label{M2N2 constraints 1}
 \frac{3}{32}&=\gamma_{\frac{1}{2},\frac{1}{2}}+\frac{3}{2} \gamma
   _{\frac{1}{4},\frac{3}{4}}\\[6pt] 
\label{M2N2 constraints 2}
\frac{5}{32}&=\frac{15}{2} \gamma _{\frac{3}{4},\frac{5}{4}}+6 \gamma
   _{\frac{1}{2},\frac{3}{2}}+\frac{7}{2} \gamma _{\frac{1}{4},\frac{7}{4}} \\[6pt]
\gamma_{\frac{1}{4},\frac{1}{4}}&=2 \left(2\gamma
   _{\frac{1}{4},\frac{1}{2}}^2+3 \gamma_{\frac{1}{4},\frac{1}{4}} \gamma
   _{\frac{1}{4},\frac{3}{4}}+\frac{5}{2} \gamma_{\frac{1}{4},\frac{5}{4}}\right)\\[6pt]
\gamma_{\frac{1}{4},\frac{1}{2}}&=\frac{4}{3} \left(2\gamma _{\frac{1}{4},\frac{1}{2}} \gamma
   _{\frac{1}{2},\frac{1}{2}}+\frac{3}{2} \gamma _{\frac{1}{4},\frac{1}{2}} \gamma
   _{\frac{1}{4},\frac{3}{4}}+\frac{3}{2} \gamma _{\frac{1}{4},\frac{1}{4}} \gamma
   _{\frac{1}{2},\frac{3}{4}}+\frac{5}{4} \gamma _{\frac{1}{2},\frac{5}{4}}+\frac{3}{2} \gamma
   _{\frac{1}{4},\frac{3}{2}}\right)\\[6pt]
\gamma_{\frac{1}{2},\frac{1}{2}}&=2\gamma
   _{\frac{1}{2},\frac{1}{2}}^2+3 \gamma _{\frac{1}{4},\frac{1}{2}} \gamma
   _{\frac{1}{2},\frac{3}{4}}+3 \gamma _{\frac{1}{2},\frac{3}{2}}\\[6pt]
\gamma _{\frac{1}{4},\frac{3}{4}}&=\frac{3}{2} \gamma _{\frac{1}{4},\frac{3}{4}}^2+2\gamma _{\frac{1}{4},\frac{1}{2}} \gamma
   _{\frac{1}{2},\frac{3}{4}}+\frac{3}{2} \gamma _{\frac{1}{4},\frac{1}{4}} \gamma
   _{\frac{3}{4},\frac{3}{4}}+\frac{5}{4} \gamma _{\frac{3}{4},\frac{5}{4}}+\frac{7}{4} \gamma
   _{\frac{1}{4},\frac{7}{4}}
\end{align}

By using the recursion relations, all 24 variables can be determined using only 6 variables
\be
f^{(1)}_{\frac{1}{2},\frac{1}{4}},\ f^{(1)}_{\frac{1}{2},\frac{1}{2}},\ f^{(1)}_{\frac{1}{2},\frac{3}{4}},\ f^{(2)}_{\frac{1}{2},\frac{1}{4}},\ f^{(2)}_{\frac{1}{2},\frac{1}{2}},\ f^{(2)}_{\frac{1}{2},\frac{3}{4}}
\ee
We have 6 constraints to determine them. 
Inserting the recursion relations (\ref{M2N2 recursion relation 1} -- \ref{M2N2 recursion relation 18}) into these constraints, we again find that 4 of the constraints are trivial and 2 of them, (\ref{M2N2 constraints 1}), (\ref{M2N2 constraints 2}), are not. The nontrivial constraints are
\bea
\frac{3}{2}&=&-4 (f^{(1)}_{\frac{1}{2},\frac{1}{2}})^2-6 f^{(1)}_{\frac{1}{2},\frac{1}{4}} f^{(1)}_{\frac{1}{2},\frac{3}{4}}-4
   (f^{(2)}_{\frac{1}{2},\frac{1}{2}})^2-6 f^{(2)}_{\frac{1}{2},\frac{1}{4}} f^{(2)}_{\frac{1}{2},\frac{3}{4}}+5,\\ 
 5 &=& -4 (f^{(1)}_{\frac{1}{2},\frac{3}{4}})^2 \big(8 (f^{(1)}_{\frac{1}{2},\frac{1}{4}})^2+5 (f^{(2)}_{\frac{1}{2},\frac{1}{4}})^2\big)-8f^{(1)}_{\frac{1}{2},\frac{3}{4}} \Big(4 f^{(1)}_{\frac{1}{2},\frac{1}{2}} f^{(2)}_{\frac{1}{2},\frac{1}{4}} f^{(2)}_{\frac{1}{2},\frac{1}{2}}+3
   f^{(1)}_{\frac{1}{2},\frac{1}{4}} f^{(2)}_{\frac{1}{2},\frac{1}{4}} f^{(2)}_{\frac{1}{2},\frac{3}{4}}\nn 
   &&+4 f^{(1)}_{\frac{1}{2},\frac{1}{4}}
   (f^{(1)}_{\frac{1}{2},\frac{1}{2}})^2+f^{(1)}_{\frac{1}{2},\frac{1}{4}}\Big) - 4 (f^{(2)}_{\frac{1}{2},\frac{3}{4}})^2 \big(5(f^{(1)}_{\frac{1}{2},\frac{1}{4}})^2+8 (f^{(2)}_{\frac{1}{2},\frac{1}{4}})^2\big)\nn 
   &&-8 \big(2 (f^{(1)}_{\frac{1}{2},\frac{1}{2}})^2
   (f^{(2)}_{\frac{1}{2},\frac{1}{2}})^2+(f^{(1)}_{\frac{1}{2},\frac{1}{2}})^4+(f^{(1)}_{\frac{1}{2},\frac{1}{2}})^2+(f^{(2)}_{\frac{1}{2},\frac{1}{2}})^4\big)
   \nn 
   &&-8f^{(2)}_{\frac{1}{2},\frac{3}{4}} \big(4 f^{(1)}_{\frac{1}{2},\frac{1}{4}} f^{(1)}_{\frac{1}{2},\frac{1}{2}} f^{(2)}_{\frac{1}{2},\frac{1}{2}}+4
   f^{(2)}_{\frac{1}{2},\frac{1}{4}} (f^{(2)}_{\frac{1}{2},\frac{1}{2}})^2+f^{(2)}_{\frac{1}{2},\frac{1}{4}}\big)-8(f^{(2)}_{\frac{1}{2},\frac{1}{2}})^2+15
\eea
We can solve these two constraints in terms of only four variables
\be\label{M2N2 variables}
f^{(1)}_{\frac{1}{2},\frac{1}{4}},\ f^{(1)}_{\frac{1}{2},\frac{1}{2}},\
f^{(2)}_{\frac{1}{2},\frac{1}{4}},\ f^{(2)}_{\frac{1}{2},\frac{1}{2}}
\ee
as

\begin{align}\label{fcopy1_1o23ov4}
f^{(1)}_{\frac{1}{2},\frac{3}{4}}&= {1\over60
   \big(\big(f_{\frac{1}{2},\frac{1}{4}}^{(1)}\big){}^2+\big(f_{\frac{1}{2},\frac{1}{4}}^{(2)}\big){}^2\big){}^2}\bigg[-2 \sqrt{2} \bigg(-2192 \big(f_{\frac{1}{2},\frac{1}{2}}^{(1)}\big){}^2
   \big(f_{\frac{1}{2},\frac{1}{2}}^{(2)}\big){}^2 \big(f_{\frac{1}{2},\frac{1}{4}}^{(1)}\big){}^2
   \big(f_{\frac{1}{2},\frac{1}{4}}^{(2)}\big){}^4\nn
   &-232 \big(f_{\frac{1}{2},\frac{1}{2}}^{(1)}\big){}^2
   \big(f_{\frac{1}{2},\frac{1}{2}}^{(2)}\big){}^2 \big(f_{\frac{1}{2},\frac{1}{4}}^{(1)}\big){}^4
   \big(f_{\frac{1}{2},\frac{1}{4}}^{(2)}\big){}^2-232 \big(f_{\frac{1}{2},\frac{1}{2}}^{(1)}\big){}^2
   \big(f_{\frac{1}{2},\frac{1}{2}}^{(2)}\big){}^2 \big(f_{\frac{1}{2},\frac{1}{4}}^{(2)}\big){}^6\nn 
   &+384
   \big(f_{\frac{1}{2},\frac{1}{2}}^{(1)}\big){}^3 \big(f_{\frac{1}{2},\frac{1}{2}}^{(2)}\big)
   \big(f_{\frac{1}{2},\frac{1}{4}}^{(1)}\big){}^3 \big(f_{\frac{1}{2},\frac{1}{4}}^{(2)}\big){}^3+1536
   \big(f_{\frac{1}{2},\frac{1}{2}}^{(1)}\big){}^3 \big(f_{\frac{1}{2},\frac{1}{2}}^{(2)}\big)
   \big(f_{\frac{1}{2},\frac{1}{4}}^{(1)}\big) \big(f_{\frac{1}{2},\frac{1}{4}}^{(2)}\big){}^5\nn 
   &-232
   \big(f_{\frac{1}{2},\frac{1}{2}}^{(1)}\big){}^4 \big(f_{\frac{1}{2},\frac{1}{4}}^{(1)}\big){}^2
   \big(f_{\frac{1}{2},\frac{1}{4}}^{(2)}\big){}^4+580 \big(f_{\frac{1}{2},\frac{1}{2}}^{(1)}\big){}^2
   \big(f_{\frac{1}{2},\frac{1}{4}}^{(1)}\big){}^2 \big(f_{\frac{1}{2},\frac{1}{4}}^{(2)}\big){}^4-20
   \big(f_{\frac{1}{2},\frac{1}{2}}^{(1)}\big){}^4 \big(f_{\frac{1}{2},\frac{1}{4}}^{(1)}\big){}^4
   \big(f_{\frac{1}{2},\frac{1}{4}}^{(2)}\big){}^2\nn 
   &+80 \big(f_{\frac{1}{2},\frac{1}{2}}^{(1)}\big){}^2
   \big(f_{\frac{1}{2},\frac{1}{4}}^{(1)}\big){}^4 \big(f_{\frac{1}{2},\frac{1}{4}}^{(2)}\big){}^2-500
   \big(f_{\frac{1}{2},\frac{1}{2}}^{(1)}\big){}^4 \big(f_{\frac{1}{2},\frac{1}{4}}^{(2)}\big){}^6+500
   \big(f_{\frac{1}{2},\frac{1}{2}}^{(1)}\big){}^2 \big(f_{\frac{1}{2},\frac{1}{4}}^{(2)}\big){}^6\nn 
   &+1536
   \big(f_{\frac{1}{2},\frac{1}{2}}^{(1)}\big) \big(f_{\frac{1}{2},\frac{1}{2}}^{(2)}\big){}^3
   \big(f_{\frac{1}{2},\frac{1}{4}}^{(1)}\big){}^3 \big(f_{\frac{1}{2},\frac{1}{4}}^{(2)}\big){}^3 +384
   \big(f_{\frac{1}{2},\frac{1}{2}}^{(1)}\big) \big(f_{\frac{1}{2},\frac{1}{2}}^{(2)}\big){}^3
   \big(f_{\frac{1}{2},\frac{1}{4}}^{(1)}\big) \big(f_{\frac{1}{2},\frac{1}{4}}^{(2)}\big){}^5\nn 
   &-840
   \big(f_{\frac{1}{2},\frac{1}{2}}^{(1)}\big) \big(f_{\frac{1}{2},\frac{1}{2}}^{(2)}\big)
   \big(f_{\frac{1}{2},\frac{1}{4}}^{(1)}\big){}^3 \big(f_{\frac{1}{2},\frac{1}{4}}^{(2)}\big){}^3-840
   \big(f_{\frac{1}{2},\frac{1}{2}}^{(1)}\big) \big(f_{\frac{1}{2},\frac{1}{2}}^{(2)}\big) \big(f_{\frac{1}{2},\frac{1}{4}}^{(1)}\big)
   \big(f_{\frac{1}{2},\frac{1}{4}}^{(2)}\big){}^5\nn 
   &-232 \big(f_{\frac{1}{2},\frac{1}{2}}^{(2)}\big){}^4
   \big(f_{\frac{1}{2},\frac{1}{4}}^{(1)}\big){}^2 \big(f_{\frac{1}{2},\frac{1}{4}}^{(2)}\big){}^4+580
   \big(f_{\frac{1}{2},\frac{1}{2}}^{(2)}\big){}^2 \big(f_{\frac{1}{2},\frac{1}{4}}^{(1)}\big){}^2
   \big(f_{\frac{1}{2},\frac{1}{4}}^{(2)}\big){}^4-500 \big(f_{\frac{1}{2},\frac{1}{2}}^{(2)}\big){}^4
   \big(f_{\frac{1}{2},\frac{1}{4}}^{(1)}\big){}^4 \big(f_{\frac{1}{2},\frac{1}{4}}^{(2)}\big){}^2\nn 
   &+500
   \big(f_{\frac{1}{2},\frac{1}{2}}^{(2)}\big){}^2 \big(f_{\frac{1}{2},\frac{1}{4}}^{(1)}\big){}^4
   \big(f_{\frac{1}{2},\frac{1}{4}}^{(2)}\big){}^2-20 \big(f_{\frac{1}{2},\frac{1}{2}}^{(2)}\big){}^4
   \big(f_{\frac{1}{2},\frac{1}{4}}^{(2)}\big){}^6+80 \big(f_{\frac{1}{2},\frac{1}{2}}^{(2)}\big){}^2
   \big(f_{\frac{1}{2},\frac{1}{4}}^{(2)}\big){}^6\nn 
   &- 250 \big(f_{\frac{1}{2},\frac{1}{4}}^{(1)}\big){}^2
   \big(f_{\frac{1}{2},\frac{1}{4}}^{(2)}\big){}^4-125 \big(f_{\frac{1}{2},\frac{1}{4}}^{(1)}\big){}^4
   \big(f_{\frac{1}{2},\frac{1}{4}}^{(2)}\big){}^2-125 \big(f_{\frac{1}{2},\frac{1}{4}}^{(2)}\big){}^6\bigg)^{1\over2}\nn 
   &-40\big(f_{\frac{1}{2},\frac{1}{2}}^{(1)}\big){}^2 \big(f_{\frac{1}{2},\frac{1}{4}}^{(1)}\big){}^3-88
   \big(f_{\frac{1}{2},\frac{1}{2}}^{(1)}\big){}^2 \big(f_{\frac{1}{2},\frac{1}{4}}^{(1)}\big)
   \big(f_{\frac{1}{2},\frac{1}{4}}^{(2)}\big){}^2+48 \big(f_{\frac{1}{2},\frac{1}{2}}^{(1)}\big)
   \big(f_{\frac{1}{2},\frac{1}{2}}^{(2)}\big) \big(f_{\frac{1}{2},\frac{1}{4}}^{(1)}\big){}^2
   \big(f_{\frac{1}{2},\frac{1}{4}}^{(2)}\big)\nn 
   &-48 \big(f_{\frac{1}{2},\frac{1}{2}}^{(1)}\big)
   \big(f_{\frac{1}{2},\frac{1}{2}}^{(2)}\big) \big(f_{\frac{1}{2},\frac{1}{4}}^{(2)}\big){}^3 -40
   \big(f_{\frac{1}{2},\frac{1}{2}}^{(2)}\big){}^2 \big(f_{\frac{1}{2},\frac{1}{4}}^{(1)}\big){}^3+8
   \big(f_{\frac{1}{2},\frac{1}{2}}^{(2)}\big){}^2 \big(f_{\frac{1}{2},\frac{1}{4}}^{(1)}\big)
   \big(f_{\frac{1}{2},\frac{1}{4}}^{(2)}\big){}^2+35 \big(f_{\frac{1}{2},\frac{1}{4}}^{(1)}\big){}^3\nn 
   &+35
   \big(f_{\frac{1}{2},\frac{1}{4}}^{(1)}\big) \big(f_{\frac{1}{2},\frac{1}{4}}^{(2)}\big){}^2\bigg]
   \\
   \nn\label{fcopy2_1o23ov4}
   f^{(2)}_{\frac{1}{2},\frac{3}{4}}&= {1\over60 \big(f_{\frac{1}{2},\frac{1}{4}}^{(2)}\big)
   \big(\big(f_{\frac{1}{2},\frac{1}{4}}^{(1)}\big){}^2+\big(f_{\frac{1}{2},\frac{1}{4}}^{(2)}\big){}^2\big){}^2}\nn 
   &\bigg[2 \sqrt{2} \big(f_{\frac{1}{2},\frac{1}{4}}^{(1)}\big) \bigg(-2192 \big(f_{\frac{1}{2},\frac{1}{2}}^{(1)}\big){}^2
   \big(f_{\frac{1}{2},\frac{1}{2}}^{(2)}\big){}^2 \big(f_{\frac{1}{2},\frac{1}{4}}^{(1)}\big){}^2
   \big(f_{\frac{1}{2},\frac{1}{4}}^{(2)}\big){}^4
   \nn 
   &-232 \big(f_{\frac{1}{2},\frac{1}{2}}^{(1)}\big){}^2
   \big(f_{\frac{1}{2},\frac{1}{2}}^{(2)}\big){}^2 \big(f_{\frac{1}{2},\frac{1}{4}}^{(1)}\big){}^4
   \big(f_{\frac{1}{2},\frac{1}{4}}^{(2)}\big){}^2-232 \big(f_{\frac{1}{2},\frac{1}{2}}^{(1)}\big){}^2
   \big(f_{\frac{1}{2},\frac{1}{2}}^{(2)}\big){}^2 \big(f_{\frac{1}{2},\frac{1}{4}}^{(2)}\big){}^6\nn 
   &+384
   \big(f_{\frac{1}{2},\frac{1}{2}}^{(1)}\big){}^3 \big(f_{\frac{1}{2},\frac{1}{2}}^{(2)}\big)
   \big(f_{\frac{1}{2},\frac{1}{4}}^{(1)}\big){}^3 \big(f_{\frac{1}{2},\frac{1}{4}}^{(2)}\big){}^3+1536
   \big(f_{\frac{1}{2},\frac{1}{2}}^{(1)}\big){}^3 \big(f_{\frac{1}{2},\frac{1}{2}}^{(2)}\big)
   \big(f_{\frac{1}{2},\frac{1}{4}}^{(1)}\big) \big(f_{\frac{1}{2},\frac{1}{4}}^{(2)}\big){}^5\nn 
   &-232
   \big(f_{\frac{1}{2},\frac{1}{2}}^{(1)}\big){}^4 \big(f_{\frac{1}{2},\frac{1}{4}}^{(1)}\big){}^2
   \big(f_{\frac{1}{2},\frac{1}{4}}^{(2)}\big){}^4+580 \big(f_{\frac{1}{2},\frac{1}{2}}^{(1)}\big){}^2
   \big(f_{\frac{1}{2},\frac{1}{4}}^{(1)}\big){}^2 \big(f_{\frac{1}{2},\frac{1}{4}}^{(2)}\big){}^4-20
   \big(f_{\frac{1}{2},\frac{1}{2}}^{(1)}\big){}^4 \big(f_{\frac{1}{2},\frac{1}{4}}^{(1)}\big){}^4
   \big(f_{\frac{1}{2},\frac{1}{4}}^{(2)}\big){}^2\nn 
   &+80 \big(f_{\frac{1}{2},\frac{1}{2}}^{(1)}\big){}^2
   \big(f_{\frac{1}{2},\frac{1}{4}}^{(1)}\big){}^4 \big(f_{\frac{1}{2},\frac{1}{4}}^{(2)}\big){}^2-500
   \big(f_{\frac{1}{2},\frac{1}{2}}^{(1)}\big){}^4 \big(f_{\frac{1}{2},\frac{1}{4}}^{(2)}\big){}^6+500
   \big(f_{\frac{1}{2},\frac{1}{2}}^{(1)}\big){}^2 \big(f_{\frac{1}{2},\frac{1}{4}}^{(2)}\big){}^6\nn 
   &+1536
   \big(f_{\frac{1}{2},\frac{1}{2}}^{(1)}\big) \big(f_{\frac{1}{2},\frac{1}{2}}^{(2)}\big){}^3
   \big(f_{\frac{1}{2},\frac{1}{4}}^{(1)}\big){}^3 \big(f_{\frac{1}{2},\frac{1}{4}}^{(2)}\big){}^3+384
   \big(f_{\frac{1}{2},\frac{1}{2}}^{(1)}\big) \big(f_{\frac{1}{2},\frac{1}{2}}^{(2)}\big){}^3
   \big(f_{\frac{1}{2},\frac{1}{4}}^{(1)}\big) \big(f_{\frac{1}{2},\frac{1}{4}}^{(2)}\big){}^5\nn 
   &-840
   \big(f_{\frac{1}{2},\frac{1}{2}}^{(1)}\big) \big(f_{\frac{1}{2},\frac{1}{2}}^{(2)}\big)
   \big(f_{\frac{1}{2},\frac{1}{4}}^{(1)}\big){}^3 \big(f_{\frac{1}{2},\frac{1}{4}}^{(2)}\big){}^3-840
   \big(f_{\frac{1}{2},\frac{1}{2}}^{(1)}\big) \big(f_{\frac{1}{2},\frac{1}{2}}^{(2)}\big) \big(f_{\frac{1}{2},\frac{1}{4}}^{(1)}\big)
   \big(f_{\frac{1}{2},\frac{1}{4}}^{(2)}\big){}^5\nn 
   &-232 \big(f_{\frac{1}{2},\frac{1}{2}}^{(2)}\big){}^4
   \big(f_{\frac{1}{2},\frac{1}{4}}^{(1)}\big){}^2 \big(f_{\frac{1}{2},\frac{1}{4}}^{(2)}\big){}^4+580
   \big(f_{\frac{1}{2},\frac{1}{2}}^{(2)}\big){}^2 \big(f_{\frac{1}{2},\frac{1}{4}}^{(1)}\big){}^2
   \big(f_{\frac{1}{2},\frac{1}{4}}^{(2)}\big){}^4-500 \big(f_{\frac{1}{2},\frac{1}{2}}^{(2)}\big){}^4
   \big(f_{\frac{1}{2},\frac{1}{4}}^{(1)}\big){}^4 \big(f_{\frac{1}{2},\frac{1}{4}}^{(2)}\big){}^2\nn 
   &+500
   \big(f_{\frac{1}{2},\frac{1}{2}}^{(2)}\big){}^2 \big(f_{\frac{1}{2},\frac{1}{4}}^{(1)}\big){}^4
   \big(f_{\frac{1}{2},\frac{1}{4}}^{(2)}\big){}^2-20 \big(f_{\frac{1}{2},\frac{1}{2}}^{(2)}\big){}^4
   \big(f_{\frac{1}{2},\frac{1}{4}}^{(2)}\big){}^6+80 \big(f_{\frac{1}{2},\frac{1}{2}}^{(2)}\big){}^2
   \big(f_{\frac{1}{2},\frac{1}{4}}^{(2)}\big){}^6\nn
   &-250 \big(f_{\frac{1}{2},\frac{1}{4}}^{(1)}\big){}^2
   \big(f_{\frac{1}{2},\frac{1}{4}}^{(2)}\big){}^4
   -125 \big(f_{\frac{1}{2},\frac{1}{4}}^{(1)}\big){}^4
   \big(f_{\frac{1}{2},\frac{1}{4}}^{(2)}\big){}^2-125 \big(f_{\frac{1}{2},\frac{1}{4}}^{(2)}\big){}^6\bigg)^{1\over2}\nn
   &+8\big(f_{\frac{1}{2},\frac{1}{2}}^{(1)}\big){}^2 \big(f_{\frac{1}{2},\frac{1}{4}}^{(1)}\big){}^2
   \big(f_{\frac{1}{2},\frac{1}{4}}^{(2)}\big){}^2-40 \big(f_{\frac{1}{2},\frac{1}{2}}^{(1)}\big){}^2
   \big(f_{\frac{1}{2},\frac{1}{4}}^{(2)}\big){}^4-48 \big(f_{\frac{1}{2},\frac{1}{2}}^{(1)}\big)
   \big(f_{\frac{1}{2},\frac{1}{2}}^{(2)}\big) \big(f_{\frac{1}{2},\frac{1}{4}}^{(1)}\big){}^3
   \big(f_{\frac{1}{2},\frac{1}{4}}^{(2)}\big)\nn 
   &+48 \big(f_{\frac{1}{2},\frac{1}{2}}^{(1)}\big)
   \big(f_{\frac{1}{2},\frac{1}{2}}^{(2)}\big) \big(f_{\frac{1}{2},\frac{1}{4}}^{(1)}\big)
   \big(f_{\frac{1}{2},\frac{1}{4}}^{(2)}\big){}^3-88 \big(f_{\frac{1}{2},\frac{1}{2}}^{(2)}\big){}^2
   \big(f_{\frac{1}{2},\frac{1}{4}}^{(1)}\big){}^2 \big(f_{\frac{1}{2},\frac{1}{4}}^{(2)}\big){}^2-40
   \big(f_{\frac{1}{2},\frac{1}{2}}^{(2)}\big){}^2 \big(f_{\frac{1}{2},\frac{1}{4}}^{(2)}\big){}^4\nn 
   &+35
   \big(f_{\frac{1}{2},\frac{1}{4}}^{(1)}\big){}^2 \big(f_{\frac{1}{2},\frac{1}{4}}^{(2)}\big){}^2+35
   \big(f_{\frac{1}{2},\frac{1}{4}}^{(2)}\big){}^4\bigg]
\end{align}    
Using these and the recursion relations (\ref{M2N2 recursion relation 1} -- \ref{M2N2 recursion relation 18}), all 24 variables can be determined in terms of only $4$ variables (\ref{M2N2 variables}). Using (\ref{f1}) and (\ref{f2}) the value of these four variables are
\begin{align} \label{input M2N2}
f^{(1)}_{{1\over2},{1\over4}}&=-{i\over2}&f^{(2)}_{{1\over2},{1\over4}}&= {i\over2}\nn
f^{(1)}_{{1\over2},{1\over2}}&= {1\over2}&f^{(2)}_{{1\over2},{1\over4}}&= {1\over2}
\end{align}
Therefore, we find the following values for the remaining variables
\begin{align}
\gamma_{\frac{1}{4},\frac{1}{4}}&=\frac{1}{4}, &\quad \gamma_{\frac{1}{4},\frac{1}{2}}&=0,&\quad\gamma_{\frac{1}{4},\frac{3}{4}}&=\frac{1}{16},&\quad\gamma_{\frac{1}{2},\frac{1}{2}}&=0,&\quad\gamma_{\frac{1}{2},\frac{3}{4}}&=0,
\nn
\gamma_{\frac{1}{2},\frac{5}{4}}&=0,&\quad \gamma_{\frac{3}{4},\frac{3}{4}}&=\frac{1}{48},&\quad
\gamma_{\frac{1}{4},\frac{5}{4}}&=\frac{1}{32},&\quad\gamma_{\frac{1}{4},\frac{3}{2}}&= 0,&\quad\gamma_{\frac{1}{4},\frac{7}{4}}&=\frac{5}{256},\nn 
\gamma_{\frac{1}{2},\frac{3}{2}}&=0,&\quad \gamma_{\frac{3}{4},\frac{5}{4}}&=\frac{3}{256},
&\quad f^{(1)}_{\frac{1}{2},\frac{3}{4}}&=\frac{i}{4},&\quad f^{(1)}_{\frac{1}{2},\frac{5}{4}}&=\frac{i}{16},\quad &f^{(1)}_{\frac{1}{2},\frac{3}{2}}&=0,
\nn 
f^{(1)}_{\frac{1}{2},\frac{7}{4}}&=\frac{i}{32},&\quad f^{(2)}_{\frac{1}{2},\frac{3}{4}}&=-\frac{i}{4},&\quad f^{(2)}_{\frac{1}{2},\frac{5}{4}}&=-\frac{i}{16},&\quad f^{(2)}_{\frac{1}{2},\frac{3}{2}}&= 0,&\quad f^{(2)}_{\frac{1}{2},\frac{7}{4}}&=-\frac{i}{32}
\end{align}
These values agree with the values that one obtains directly from (\ref{MN gamma}), (\ref{f1}) and (\ref{f2}). We note that the number of inputs is 4, in agreement with equation (\ref{N inputs}) for $(M,N)=(2,2)$ and $N^{\text{constr}}=2$.

In the above scenario, we note also that quantities whose final state energies are multiples of $1/2$ vanish (except for $f^{(i)}_{{1\over2},{1\over2}}$). This is because both initial copies have winding $2$, respectively, and the final copy winding $4$, suggesting that all mode numbers, both initial and final, can be rescaled by a factor of $2$. After this rescaling, the initial copies then have winding $1$ and the final copy has winding $2$. This case was already studied in \cite{Avery:2010er,Avery:2010hs}. As shown in (\ref{f n0}) and (\ref{g n0}), the $\gamma$'s were argued to vanish when the mode numbers were integers and the $f$'s we argued to vanish when the final mode numbers were integers (unless the final energy was equal to the initial energy).  

\section{Discussion}\label{section 10}

In this paper we extended the techniques developed in \cite{Guo:2022sos, Guo:2022zpn} to bootstrap the effects of the twist operator $\sigma_2$ for multiwound initial states. The major difference between this scenario and those in \cite{Guo:2022sos, Guo:2022zpn} is that the coefficients of pair creation and propagation are coupled together in the equations. This arises from the fact that $L_{-1}$ acting on singly wound initial states gives zero whereas $L_{-1}$ acting on multiwound initial states does not, which is due to the presence of fractional modes in the multiwound sectors. This significantly increased the work required in finding solutions as opposed to the case which considered only singly wound copies in the initial state. 

We investigated the scenario beginning with initial states containing two copies where copy 1 had winding $M$ and copy 2 had winding $N$. We looked at the effects produced by twisting together initial states in these winding sectors into final states living on a copy of winding $M+N$. Using a set of Virasoro generators $L_{-1},L_0,L_1,L_{n>1}$ we derived general relations for the effects of the twist operator. Because we were considering multiwound initial states, the nonvanishing of $L_{-1}$ on the multiwound vacua gave rise to relations which coupled together pair creation, $\gamma$, with propagation, $f$. 

To solve these coupled equations systematically, we organized them into recursion relations and constraints in the following way. We first used the relation derived from $L_0$ to determine the dependence of the coordinate $w$. Next, we organized the $L_{-1}$ relation for pair creation and the $L_1$ relation for propagation as recursion relations. By using these two classes of equations, we can effectively determine the infinite number of coefficients associated with pair creation and propagation, using only a finite number of inputs. These inputs are the low-energy propagations $f^{(1)}_{\frac{r}{M}<1,\frac{p}{M+N}<1}$ and $f^{(2)}_{\frac{r}{N}<1,\frac{p}{M+N}<1}$, which have initial and final energies less than 1.
In addition, there are two types of constraints to consider. 
Firstly, we have constraint (\ref{c1}) derived from $L_{n>0}$ for pair creation from terms without any modes. Secondly, we have relations (\ref{c2}) derived from $L_1$ for pair creation from terms with two modes. Each type of constraint is infinite in number, while only a finite number of inputs need to be solved. 

Due to the complication of these constraints, it was unclear if we could use them to determine all the inputs. To investigate this, we studied explicit examples for $(M,N)=(2,1),(3,1),(2,2)$ and considered low energy constraints from the aforementioned two types. We found that the second type of constraint was always trivial. For the first type, the $L_1$ constraint was always nontrivial, while the $L_2$ constraint was only nontrivial for the case $(M,N)=(3,1),(2,2)$. We used these constraints to reduce the number of inputs.
It remains unclear whether the two types of constraints are sufficient to determine all the inputs, given that we have only considered low-energy constraints within each type. Based on the explicit examples, we conjecture that as $M$ and $N$ increase, more constraints from the first type will become nontrivial, and the second type of constraint will always be trivial for any $M$ and $N$. Even though not all the inputs can be determined, finding the infinite number of coefficients using only a finite number of inputs is still a nontrivial step.

There are potential directions for improving the constraints. One possibility is to continue considering constraints of the same type but at higher energies, which we have not explored in the examples investigated here. As we reach higher energies, there may be more nontrivial constraints. Another approach is to consider additional types of constraints. For instance, we could study terms involving four or more modes in (\ref{L1 start}), and likewise for other $L_n$. We could also introduce more initial modes, such as including two initial modes in (\ref{L1 start}). This approach would further couple the contraction with pair creation and propagation. Although it would introduce more coefficients to solve, it would also generate more constraints. It's also possible that one may need to find relations using fractional Virasoro generators \cite{Burrington:2022dii,Burrington:2022rtr}. In this case, one would need to know how to move these generators through the twist operator to obtain relations in the same manner we use in this paper. These approaches have the potential to determine all the required inputs. 

Finally, an overall primary goal to which these computations contribute, is to compute the effects when an arbitrary number of twist operators are inserted. This is because in certain scenarios, particularly in the D1D5 CFT, these twist operators play a fundamental role in marginal deformations away from free theories towards supergravity theories. By understanding these twist effects, one can better understand how physical processes studied at the free point actually map to the supergravity point. 
While the current work only considers a single twist, we believe these methods, along with those of \cite{Guo:2022sos,Guo:2022zpn}, in an appropriate setting, can be combined and adapted to understand the effects produced by multiple twists. We plan to return to this in future work.

\section*{Acknowledgements}

The work of B.G. and S.D.H is supported by ERC Grant787320 - QBH Structure.

\bibliographystyle{JHEP}
\bibliography{bibliography.bib}

\end{document}